\documentclass[11pt,a4paper]{article} 
\pdfoutput=1
\usepackage{jcappub}
\usepackage{graphicx}
\usepackage[lofdepth,lotdepth]{subfig}
\usepackage{dcolumn}
\usepackage{bm}
\usepackage{epsfig}
\usepackage{color}
\usepackage{rotating}
\usepackage{psfrag}

\title{Exploring $\nu_\tau - \nu_s$ mixing with cascade events in DeepCore}

\author[a]{Arman Esmaili}
\author[b]{, Francis Halzen}
\author[a,c]{, O. L. G. Peres}
\emailAdd{aesmaili@ifi.unicamp.br}
\emailAdd{halzen@icecube.wisc.edu}
\emailAdd{orlando@ifi.unicamp.br}
\affiliation[a]{Instituto de Fisica Gleb Wataghin - UNICAMP, 13083-859, Campinas, SP, Brazil}
\affiliation[b]{Wisconsin IceCube Particle Astrophysics Center and Department of Physics, University of Wisconsin, Madison, WI 53706, USA}
\affiliation[c]{Abdus Salam International Centre for Theoretical Physics, ICTP, I-34010, Trieste, Italy}

\abstract{The atmospheric neutrino data collected by the IceCube experiment and its low-energy extension DeepCore provide a unique opportunity to probe the neutrino sector of the Standard Model. In the low energy range the experiment have observed neutrino oscillations, and the high energy data are especially sensitive to signatures of new physics in the neutrino sector. In this context, we previously demonstrated the unmatched potential of the experiment to reveal the existence of light sterile neutrinos. The studies are routinely performed in the simplest $3+1$ model concentrating on disappearance of muon neutrinos of TeV energy as a result of their mixing with a sterile neutrino. We here extend this analysis to include cascade events that are secondary electromagnetic and hadronic showers produced by neutrinos of all flavors. We find that it is possible to probe the complete parameter space of $3+1$ model, including the poorly constrained mixing of the sterile neutrino to tau neutrinos. We show that $\nu_\tau-\nu_s$ mixing results into a unique signature in the data that will allow IceCube to obtain constraints well below the current upper limits.}

\begin{document}

\maketitle

\section{\label{sec:intro}Introduction}

Oscillation of neutrino flavor has been confirmed by a variety of data from atmospheric, solar, reactor and accelerator neutrino experiments~\cite{GonzalezGarcia:2007ib}. A consistent description of the oscillation mechanism, that accommodates all the data successfully, can be achieved by assuming three neutrino mass eigenstates (with at least two nonzero masses) and the mixing between them. In this $3\nu$ framework all the neutrino states have the standard weak interaction and the oscillation of neutrino flavor results from the differences between mass and weak interaction eigenstates. Although the majority of experimental data is consistent with the $3\nu$ framework, there are anomalies that challenge this framework. These anomalies include: the long-standing evidence for $\bar{\nu}_\mu\to\bar{\nu}_e$ oscillation in the LSND experiment~\cite{Aguilar:2001ty}, the recent excess observed in both the $\bar{\nu}_\mu\to\bar{\nu}_e$ and $\nu_\mu\to\nu_e$ channels in the MiniBooNE experiment~\cite{AguilarArevalo:2012va,Aguilar-Arevalo:2013pmq} which supports the LSND result, the $\bar{\nu}_e$ disappearance in short baseline reactor experiments associated with the recent re-evaluation of reactor neutrino flux showing a 3\% increase~\cite{Mueller:2011nm,Huber:2011wv}, and the $\nu_e$ disappearance observed in the GALLEX~\cite{Kaether:2010ag} and SAGE~\cite{Abdurashitov:2005tb} experiments (Gallium anomaly). In addition to these anomalies in neutrino oscillation experiments, cosmological data favor the existence of extra light degrees of freedom in the universe~\cite{Komatsu:2010fb,Seljak:2006bg,GonzalezGarcia:2010un,Archidiacono:2011gq,Hamann:2010bk}. For a thorough review of all these anomalies see~\cite{Abazajian:2012ys}.

A minimal framework that can partially accommodate the anomalies mentioned above is the $3+1$ model, although the consistency of this model to various data sets is a matter of debate. This model adds an almost sterile state to the $3\nu$ framework, with the mass of the new state $\sim\mathcal{O}(1)$~eV. The sterile state can mix with the three mostly active neutrino states, inducing flavor oscillation between the sterile flavor state $\nu_s$ and the three active flavors $\nu_e$, $\nu_\mu$ and $\nu_\tau$. Within this scheme, the reactor and Gallium  anomalies can be explained as $\nu_e-\nu_s$ mixing, while the LSND and MiniBooNE anomalies are the result of both $\nu_e-\nu_s$ and $\nu_\mu-\nu_s$ mixing. Considering the complete parameter space of the $3+1$ model, $\nu_s$ can also mix with $\nu_\tau$, a component of the scheme that is out of reach of the experiments mentioned above and poorly constrained by a variety of proposed experiments.

In this paper we show how to probe $\nu_\tau-\nu_s$ mixing using cascade events in IceCube. The effect of sterile neutrinos on the atmospheric neutrino flux and the possibility of its detection have been previously considered in~\cite{Nunokawa:2003ep,Choubey:2007ji,Razzaque:2012tp,Razzaque:2011ab,Barger:2011rc,Halzen:2011yq,Esmaili:2012nz}. In the presence of sterile neutrinos matter effects on neutrinos propagating through the Earth enhance active-sterile neutrino oscillations. This leads to observable distortions in energy and zenith distributions of the atmospheric neutrino flux. In this regard, neutrino telescopes such as IceCube are perfect detectors in the search for sterile neutrinos. In studies performed so far only events producing muon tracks in IceCube, that are sensitive to $\nu_\mu-\nu_s$ mixing, have been considered. In this paper we consider cascade events and we show that it is possible to probe mixing of $\nu_s$ with all active flavors, including the $\nu_\tau-\nu_s$ component whose measurement is challenging.  

The paper is organized as follows: in section~\ref{sec:prob} we review the $3+1$ model presenting the formalism describing vacuum oscillation probabilities in section~\ref{sec:vacuumprob} and matter effects in section~\ref{sec:matterprob}. In section~\ref{sec:chi2} we calculate the cascade event rates from atmospheric neutrinos in the DeepCore part of IceCube and define the $\chi^2$ function used in our analysis. In section~\ref{sec:q1424} we derive the sensitivity of DeepCore to $\nu_e-\nu_s$ and $\nu_\mu-\nu_s$ mixings, and in section~\ref{sec:q34}, which contains the main result of this paper, we present the sensitivity of DeepCore to $\nu_\tau-\nu_s$ mixing. Finally, we conclude in section~\ref{sec:conclusion}.

\section{\label{sec:prob}Framework of $3+1$ model}

The $3+1$ model consist of the three standard active neutrino states $(\nu_1,\nu_2,\nu_3)$ and a mostly sterile state $\nu_4$ with mass $m_4\sim 1$~eV. In this model, in addition to the standard mixing parameters ($\theta_{12}$, $\theta_{13}$, $\theta_{23}$, $\Delta m_{31}^2\simeq\Delta m_{32}^2$ and $\Delta m_{21}^2$), four new mixing parameters are introduced: three mixing angles $\theta_{i4}$ (where $i=1,2,3$), and one new mass-squared difference $\Delta m_{41}^2\equiv m_4^2-m_1^2\simeq\Delta m_{42}^2\simeq\Delta m_{43}^2$, assuming that CP is conserved in the neutrino sector. The master evolution equation of neutrinos in the $3+1$ scheme can be written as:

\begin{equation}\label{eq:evolution}
i\frac{{\rm d}\nu_\alpha}{{\rm d}r}=\left[ \frac{1}{2E_\nu}\mathbf{U_4} \mathbf{M^2}\mathbf{U_4}^\dagger+\mathbf{V}(r) \right]_{\alpha\beta}\nu_\beta~,
\end{equation}  
where $\mathbf{M^2}$ is the $4\times4$ mass-squared differences matrix given by
\begin{equation}
\mathbf{M^2}=\mathbf{{\rm diag}}\left(0,\Delta m_{21}^2, \Delta m_{31}^2, \Delta m_{41}^2\right)~,
\end{equation}
and $\mathbf{V}(r)$ is the diagonal matrix of matter potential as a function of distance $r$, given by
\begin{equation}
\mathbf{V}(r)=\sqrt{2}G_F\mathbf{{\rm diag}}\left(N_e(r), 0,0, N_n(r)/2\right)~.
\end{equation}
Here $G_F$ is the Fermi constant; $N_e(r)$ and $N_n(r)$ are the electron and neutron number density profile of the medium neutrinos propagate in it. The $4\times4$ unitary mixing matrix $\mathbf{U_4}$ can be parametrized as\footnote{Since the rotation matrices $\mathbf{R^{14}}$ and $\mathbf{R^{23}}$ commute, the parametrization in eq.~(\ref{eq:para}) is equivalent to the parametrization $\mathbf{U_4}=\mathbf{R^{34}}\mathbf{R^{24}}\mathbf{R^{23}}\mathbf{R^{14}}\mathbf{R^{13}}\mathbf{R^{12}}$.}~\cite{deGouvea:2008nm}:
\begin{equation}\label{eq:para}
\mathbf{U_4}= \mathbf{R^{34}}(\theta_{34})\mathbf{R^{24}}(\theta_{24})\mathbf{R^{14}}(\theta_{14})\mathbf{R^{23}}(\theta_{23})\mathbf{R^{13}}(\theta_{13})\mathbf{R^{12}}(\theta_{12})~,
\end{equation}
where $\mathbf{R^{ij}}(\theta_{ij})$ ($i,j=1,\ldots,4$ and $i<j$) is the $4\times4$ rotation matrix in the $ij$-plane with the angle $\theta_{ij}$, with elements
\begin{equation}
\left[\mathbf{R^{ij}}(\theta_{ij})\right]_{kl}=(\delta_{ik}\delta_{il}+\delta_{jk}\delta_{jl})c_{ij}+(\delta_{ik}\delta_{jl}-\delta_{il}\delta_{jk})s_{ij}+\left[(1-\delta_{ik})(1-\delta_{jl})+(1-\delta_{il})(1-\delta_{jk})\right]\delta_{kl}/2~,
\end{equation}
where $c_{ij}\equiv\cos\theta_{ij}$ and $s_{ij}\equiv\sin\theta_{ij}$.

Various experimental data can be used to constrain the oscillation parameters  in $3+1$ model: $\theta_{14}$, $\theta_{24}$, $\theta_{34}$ and $\Delta m_{41}^2$. In this section we discuss the oscillation probabilities, obtained from the evolution equation in eq.~(\ref{eq:evolution}), and the oscillation parameters that can be constrained in each experiment. In section~\ref{sec:vacuumprob} we briefly summarize the oscillation probabilities in vacuum and the current constrains/indications on mixing parameters. In section~\ref{sec:matterprob} we study the oscillation probabilities in matter, which is relevant for this paper, for the atmospheric neutrinos propagating inside the Earth.

\subsection{\label{sec:vacuumprob}Oscillation probabilities in vacuum}

For reactor and short baseline accelerator neutrino experiments searching for the active-sterile oscillation, the oscillation probability can be calculated from eq.~(\ref{eq:evolution}) by omitting the $\mathbf{V}(r)$ term and the oscillation can be effectively approximated by a two flavors approximation. The reactor neutrino experiments searching for $\bar{\nu}_e$ disappearance, actually measure the survival probability

\begin{equation}\label{eq:probee}
P(\bar{\nu}_e\to\bar{\nu}_e)\cong 1- \sin^2 2\vartheta_{ee}\sin^2 \left( \frac{\Delta m^2_{41} L}{4E_\nu} \right)~,
\end{equation} 
where 
\begin{equation}\label{eq:qee}
\sin^2 2\vartheta_{ee}\equiv 4 |\mathbf{U}_{e4}|^2 \left( 1- |\mathbf{U}_{e4}|^2 \right)~.
\end{equation}
From the parametrization of eq.~(\ref{eq:para}) $\mathbf{U}_{e4}=\sin\theta_{14}$ and $\vartheta_{ee}\equiv\theta_{14}$. Recent re-evaluations of the reactor $\bar{\nu}_e$ flux show a $\sim 3\%$ increase in the flux~\cite{Mueller:2011nm,Huber:2011wv} hinting at a deviation of $P(\bar{\nu}_e\to\bar{\nu}_e)$ from one. The global analysis of all the reactor neutrino experiments leads to best-fit values~\cite{Giunti:2012tn}: $(\Delta m_{41}^2,\sin^2 2\vartheta_{ee})=(1.9~{\rm eV}^2,0.13)$ with a $2.8\sigma$ significance. The inclusion of the Gallium anomaly, solar and $\nu_e{\rm C}$ data leads to larger values for $\Delta m_{41}^2$ with a significance $\sim3\sigma$; see \cite{Giunti:2012tn}. Tritium beta decay experiments provide an independent way of constraining $\vartheta_{ee}$ because a sterile neutrino will distort the electron spectrum~\cite{Farzan:2001cj,Farzan:2002zq,deGouvea:2006gz,Riis:2010zm,Formaggio:2011jg,Esmaili:2012vg,Kraus:2012he}. Although the current data from the Mainz experiment are barely sensitive to the parameter space favored by the reactor anomaly~\cite{Kraus:2012he}, the forthcoming KATRIN experiment is expected to cover the complete allowed region in $(\Delta m_{41}^2,\sin^2 2\vartheta_{ee})$ space~\cite{Esmaili:2012vg}.  

The LSND~\cite{Aguilar:2001ty} and MiniBooNE~\cite{AguilarArevalo:2012va} experiments search for excess in the $\nu_\mu(\bar{\nu}_\mu)\to~\nu_e(\bar{\nu}_e)$ oscillation channels, using baselines of $L_{\rm LSND}=30$~m and $L_{\rm MiniBooNE}=540$~m. With these baselines we have $\Delta m_{31}^2L_{\rm LSND,MiniBooNE}/(4E_\nu)\ll1$ and $\Delta m_{21}^2L_{\rm LSND,MiniBooNE}/4E_\nu\ll1$, and thus the probability of $\nu_\mu(\bar{\nu}_\mu)\to~\nu_e(\bar{\nu}_e)$ oscillation can be written as:

\begin{equation}\label{eq:probemu}
P(\nu_\mu(\bar{\nu}_\mu)\to\nu_e(\bar{\nu}_e))=\sin^2 2\vartheta_{e\mu} \sin^2 \left( \frac{\Delta m^2_{41} L}{4E_\nu} \right)~,
\end{equation} 
where 
\begin{equation}
\sin^2 2\vartheta_{e\mu}\equiv 4 |\mathbf{U}_{e4}|^2 |\mathbf{U}_{\mu4}|^2~.
\end{equation}
In our parametrization $\mathbf{U}_{\mu4}=\cos\theta_{14}\sin\theta_{24}$, therefore LSND and MiniBooNE are sensitive to both $\theta_{14}$ and $\theta_{24}$. The recent MiniBooNE fit of neutrino and anti-neutrino data gives the best-fit values $(\Delta m_{41}^2,\sin^2 2\vartheta_{e\mu})=(0.037~{\rm eV}^2,1.0)$ with $3.8\sigma$ significance~\cite{AguilarArevalo:2012va,Aguilar-Arevalo:2013pmq}. However, the large best-fit value of $\sin^2 2\vartheta_{e\mu}$ mainly comes from the anti-neutrino data such that the best-fit values from the analysis of neutrino data is $(\Delta m_{41}^2,\sin^2 2\vartheta_{e\mu})=(3.14~{\rm eV}^2,0.002)$.

The MINOS experiment, with near and far detectors at baselines 1.04~km and 735~km from the source, measures the $\nu_\mu\to\nu_\mu$ oscillation probability by comparing the rate of events induced by charged current (CC) interactions at near and far detectors. Although the deficit of CC events at the far detector is consistent with $\nu_\mu\to\nu_\tau$ oscillations, in the $3+1$ model, this deficit is sensitive to active-sterile mixing. Assuming the range of $\Delta m_{41}^2$ to be such that the active-sterile oscillation do not take place at the near detector and average out at the far detector ({\it i.~e.}; $0.2\lesssim\Delta m_{41}^2 ({\rm eV}^2)\lesssim1$), the $\nu_\mu$ survival probability can be written as~\cite{Adamson:2010wi}:

\begin{equation}\label{eq:probmumu}
P(\nu_\mu\to\nu_\mu)=1-4\left\{ |\mathbf{U}_{\mu3}|^2 \left( 1-|\mathbf{U}_{\mu3}|^2-|\mathbf{U}_{\mu4}|^2 \right) \sin^2 \left( \frac{\Delta m^2_{\rm 31} L}{4E_\nu} \right) +\frac{|\mathbf{U}_{\mu4}|^2}{2} (1-|\mathbf{U}_{\mu4}|^2)  \right\}~.
\end{equation} 
Analogous to eq.~(\ref{eq:qee}), we can define
\begin{equation}\label{eq:qemu}
\sin^2 2\vartheta_{\mu\mu}\equiv 4 |\mathbf{U}_{\mu4}|^2 \left( 1- |\mathbf{U}_{\mu4}|^2 \right)~,
\end{equation}
and the measurement of probability in eq.~(\ref{eq:probmumu}) by MINOS is sensitive to $\vartheta_{\mu\mu}$ (or $\theta_{24}$, if we assume $\theta_{14}=0$ in the fit of MINOS data). In the analysis of reference~\cite{Adamson:2011ku}, the MINOS collaboration reported the upper limit $\mathbf{U}_{\mu4}\leq0.019$ at 90\% C.L., which corresponds to $\sin^2 2\vartheta_{\mu\mu}\leq0.075$. Also, in~\cite{Giunti:2011hn} the MINOS is re-analyzed for a wider range of $\Delta m_{41}^2$, yielding the same upper limit for $\Delta m_{41}^2\lesssim 1~{\rm eV}^2$ ($\sin^2 2\vartheta_{\mu\mu}\leq 0.09$) and a weaker limit for $\Delta m_{41}^2\gtrsim 1~{\rm eV}^2$. The best-fit point of~\cite{Giunti:2011hn} is $(\Delta m_{41}^2,\sin^2 2\vartheta_{\mu\mu})=(6.76~{\rm eV}^2,0.48)$.

All matrix elements appearing in the oscillation probabilities in eqs.~(\ref{eq:probee}), (\ref{eq:probemu}) and (\ref{eq:probmumu}) can be written in terms of the mixing angles $\theta_{14}$ and $\theta_{24}$.
Thus, the three effective mixing angles $(\vartheta_{ee},\vartheta_{e\mu},\vartheta_{\mu\mu})$ are not independent and satisfy the following relation:

\begin{equation}
\sin^2 2\vartheta_{e\mu}=4\sin^2 \vartheta_{ee}\sin^2 \vartheta_{\mu\mu}~,
\end{equation}    
which makes $\sin^2 2\vartheta_{e\mu}$ double suppressed and leads to some tension between the results of the above mentioned experiments (see~\cite{Giunti:2011hn,Peres:2000ic,Kopp:2011qd}). 

One of the available experimental data sensitive to the angle $\theta_{34}$ is from MINOS. By comparing the rate of events induced by neutral current interaction at near and far detectors of MINOS, it is possible to measure the $\nu_\mu\to\nu_s$ oscillation probability. In $3+1$ scenario, with the assumption of no oscillation at near detector and averaged oscillation at far detector, we have (assuming CP conservation)~\cite{Adamson:2010wi}:
\begin{equation}\label{eq:probmus}
P(\nu_\mu\to\nu_s)=4\left\{ \left( |\mathbf{U}_{\mu3}|^2 |\mathbf{U}_{s3}|^2+\mathbf{U}_{\mu3}\mathbf{U}_{\mu4}\mathbf{U}_{s3}\mathbf{U}_{s4} \right) \sin^2 \left( \frac{\Delta m^2_{\rm 31} L}{4E_\nu} \right) +\frac{|\mathbf{U}_{\mu4}|^2 |\mathbf{U}_{s4}|^2}{2} \right\}~.
\end{equation} 
As can be seen, the $\nu_\mu\to\nu_s$ oscillation probability depends on the $\mathbf{U}_{s3}$ and $\mathbf{U}_{s4}$ matrix elements, which in the parametrization of eq.~(\ref{eq:para}), are given by: 
$$\mathbf{U}_{s3}=-\cos\theta_{24}\cos\theta_{34}\sin\theta_{13}\sin\theta_{14}-\cos\theta_{13}\cos\theta_{34}\sin\theta_{23}\sin\theta_{24}-\cos\theta_{13}\cos\theta_{23}\sin\theta_{34}$$ 
and 
$$ \mathbf{U}_{s4}=\cos\theta_{14}\cos\theta_{24}\cos\theta_{34}~.$$
The dependence of these matrix elements on $\theta_{34}$ leads to the unique opportunity to constrain this angle using MINOS data. From the measurement of muon disappearance probability MINOS collaboration obtained the upper limit of $\sin^2 2\theta_{34}<0.59$ at 90\% C.L.~\cite{Adamson:2011ku,Adamson:2010wi}.

To probe the mixing angle $\theta_{34}$ by vacuum oscillation, i.e. when matter effects are subdominant, oscillation probabilities containing one of the matrix elements from the third or fourth rows of $\mathbf{U}_4$ should be measured. This implies that in order to constrain $\theta_{34}$ one has to measure the oscillation probabilities $\nu_\alpha\to\nu_\tau$ or $\nu_\alpha\to\nu_s$, where $\alpha=e,\mu,\tau$. Among these the $\nu_\mu\to\nu_s$ oscillation has been probed by MINOS (see eq.~(\ref{eq:probmus})) as discussed above, and the $\nu_\mu\to\nu_\tau$ channel can be explored by the OPERA experiment~\cite{Donini:2007yf,Migliozzi:2011bj}. Also, the sensitivity of proposed neutrino factories to $\theta_{34}$ through $\nu_e\to\nu_\tau$ and $\nu_\mu\to\nu_\tau$ channels has been studied in~\cite{Donini:2001xy,Donini:2008wz,Meloni:2010zr}.

Experiments measuring the low energy atmospheric neutrinos and solar neutrinos can also constrain $\theta_{34}$ due to matter effects induced by sterile neutrino at atmospheric and solar scale. It is shown in~\cite{Maltoni:2007zf} that a non-zero value for $\theta_{34}$ induces a effective matter potential, that changes the favored $\nu_{\mu}\to \nu_{\tau}$ oscillation in low energy atmospheric neutrino analyses. From~\cite{Maltoni:2007zf} the upper limits $\sin^22\theta_{34}\lesssim 0.75$ at 90\% C.L. can be obtained which is comparable to the MINOS limit. The matter effects induced by sterile neutrinos would also affect solar neutrinos. In~\cite{Maltoni:2001bc} it shown that from solar neutrino experiments the upper bound $|U_{s1}|^2+|U_{s2}|^2 < 0.52$ (at 99\% C.L.) can be obtained, which barely constrain $\theta_{34}$. In the next section we discuss the matter effect in the propagation of high energy atmospheric neutrinos through the Earth in the presence of sterile neutrinos.

\subsection{\label{sec:matterprob}Oscillation probabilities in matter}

The signatures of sterile neutrinos in $3+1$ model with $\Delta m_{41}^2\sim 1~{\rm eV}^2$ on high energy atmospheric neutrino fluxes measured by IceCube/DeepCore have been studied in~\cite{Nunokawa:2003ep,Choubey:2007ji,Razzaque:2012tp,Razzaque:2011ab,Barger:2011rc,Halzen:2011yq,Esmaili:2012nz}. In these works the effect of sterile neutrinos on zenith distribution of muon-track events induced by atmospheric neutrinos and its measurement by IceCube/DeepCore were analyzed. The zenith distribution of muon-track events in IceCube is sensitive to $\theta_{24}$ and stringent upper limits on $\theta_{24}$ have been obtained in~\cite{Esmaili:2012nz} from the analysis of muon-track events collected with the partially deployed detector consisting of 40 out of the final 86 strings of photomultipliers. In this paper, we study the $3+1$ model exploiting cascade events induced by atmospheric neutrinos in DeepCore. The cascade events originate from: i) neutral current interaction of all active neutrino flavors ($\nu_e,\nu_\mu,\nu_\tau$) and ($\bar{\nu}_e,\bar{\nu}_\mu,\bar{\nu}_\tau$); ii) charged current interaction of $\nu_e$ and $\bar{\nu}_e$; iii) charged current interaction of $\nu_\tau$ and $\bar{\nu}_\tau$. The atmospheric flux consist of $\nu_e$ and $\nu_\mu$ (also $\bar{\nu}_e$ and $\bar{\nu}_\mu$) neutrinos, although the electron neutrino flux is smaller than the muon neutrino flux by over one order of magnitude at TeV energy. While muon-track measurements depend on $P(\nu_\mu\to\nu_\mu)$ and $P(\nu_e\to\nu_\mu)$, the rate of cascade events probes the oscillation probabilities: $P(\nu_e\to\nu_\alpha)$ and $P(\nu_\mu\to\nu_\alpha)$, where $\alpha=e,\mu,\tau$. However, it should be emphasized that, because the flux of atmospheric $\nu_e$ and $\bar{\nu}_e$ is small, the contribution of $P(\nu_e\to\nu_\alpha)$ to the events, both muon-track and cascades, induced by atmospheric neutrinos is small. 

In the following we show the oscillograms of probabilities for various sets of values for the additional mixing parameters of $3+1$ model. The probabilities are calculated by the numerical integration of eq.~(\ref{eq:evolution}), with the standard mixing parameters: $\sin^2 \theta_{12}=0.3$, $\sin^2 \theta_{13}=0.02$, $\sin^2 \theta_{23}=0.5$, $\Delta m_{31}^2=2.4\times10^{-3}~{\rm eV}^2$ and $\Delta m_{21}^2=7.5\times10^{-5}~{\rm eV}^2$~\cite{GonzalezGarcia:2012sz}. In order to avoid conflict with cosmological constraints on the sum of neutrino masses, we assume a $3+1$ model with $\Delta m_{41}^2\geq0$. The electron and neutron number density profiles of the Earth have been taken from the PREM model~\cite{prem}. In the following we define several possibilities according to the vanishing or non-vanishing values of $\theta_{i4}$, and for each case we discuss the oscillation probability pattern of atmospheric neutrinos.

\begin{figure}[t!]
\begin{center}
\subfloat[$P(\nu_e\to\nu_e)$ for $\sin^22\theta_{14}=0.1$]{
 \includegraphics[width=0.5\textwidth]{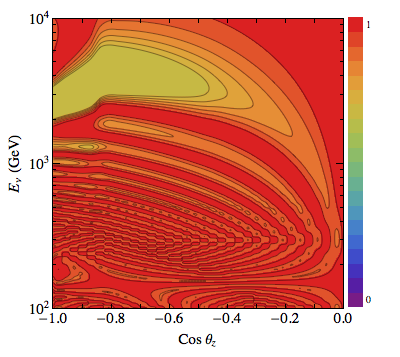}
  \label{fig:etoe21,8,1,1}
}
\subfloat[$P(\nu_e\to\nu_e)$ for $\sin^22\theta_{14}=0.3$]{
 \includegraphics[width=0.5\textwidth]{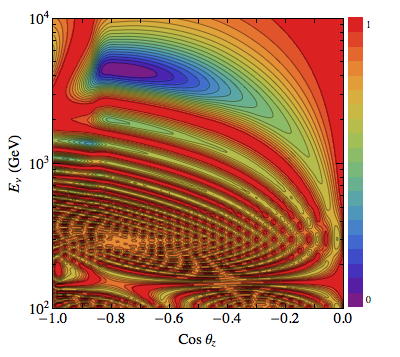}
 \label{fig:etoe,21,11,1,1}
}
\end{center}
\caption{\label{fig:probcaseI}Oscillograms of survival probability $P(\nu_e\to\nu_e)$ for $\Delta m_{41}^2=1\,{\rm eV}^2$, $\theta_{24}=\theta_{34}=0$ and different values of $\sin^22\theta_{14}$ indicated in each sub-caption.}
\end{figure}

\paragraph{Case I: $\theta_{14}\neq0$, $\theta_{24}=0$ and $\theta_{34}=0$.} In this case the $\nu_4$ ($\bar{\nu}_4$) mass eigenstate contributes to the $\nu_e$ ($\bar{\nu}_e$) flavor eigenstate only. The matter effects from the propagation of neutrino through the Earth enhance $\nu_e\to\nu_s$ oscillation. Thus, we expect enhancement of the conversion of $\nu_e$ atmospheric flux to $\nu_s$ at energies $E_\nu \sim 4-5~{\rm TeV}~(\Delta m_{41}^2/1~{\rm eV}^2)$. Due to the change in sign of matter potential for $\bar{\nu}_e$, enhancement of active-sterile oscillation takes place just for $\nu_e$. In Figure~\ref{fig:probcaseI} we show the oscillogram for electron neutrino survival probability as a function of $\cos\theta_z$ and $E_\nu$ in the range $10^2-10^4$~GeV. As can be seen, for nonzero values of $\theta_{14}$ a dip develops in the $P(\nu_e\to\nu_e)$; while in standard $3\nu$ framework $P(\nu_e\to\nu_e)=1$ in the energy range we showed.

\paragraph{Case II: $\theta_{14}=0$, $\theta_{24}\neq0$ and $\theta_{34}=0$.} In this case the nonzero contribution of $\nu_4$ ($\bar{\nu}_4$) is to $\nu_\mu$ ($\bar{\nu}_\mu$). Due to sign of the effective potential in the muon-sterile (anti-)neutrino system, the enhancement of oscillation takes place just for $\bar{\nu}_\mu$ atmospheric flux. The enhancement leads to a dip in the $\bar{\nu}_\mu$ survival probability and a peak in the $P(\bar{\nu}_\mu\to\bar{\nu}_s)$. It should be noticed that the previous analyses of $3+1$ model with IceCube/DeepCore using muon-track events~\cite{Razzaque:2012tp,Razzaque:2011ab,Barger:2011rc,Halzen:2011yq,Esmaili:2012nz}, cover this case and the limits obtained in these papers constrain $\theta_{24}$ (or $\vartheta_{\mu\mu}$). The oscillograms for this case are similar to those for \textbf{Case I} as has been shown in~\cite{Esmaili:2012nz}.

\paragraph{Case III: $\theta_{14}=0$, $\theta_{24}=0$ and $\theta_{34}\neq0$.} In this case the $\nu_4$ ($\bar{\nu}_4$) contributes to $\nu_\tau$ ($\bar{\nu}_\tau$) and we therefore expect enhancement of active-sterile oscillation for a $\bar{\nu}_\tau$ beam. However, the anticipated atmospheric $\bar{\nu}_\tau$ flux produced in decay of charmed particles is quite small in the energy range $E_\nu \lesssim 10$~TeV and has not been observed. Thus, by setting $\theta_{14}=\theta_{24}=0$, it is not possible to probe $\theta_{34}$ by atmospheric neutrino data through resonance active-sterile conversion. However, active-sterile oscillation evidence from short baseline experiments force $\theta_{14}\neq0$ and $\theta_{24}\neq0$; which we discuss next.

\paragraph{Case IV: $\theta_{14}\neq0$, $\theta_{24}\neq0$ and $\theta_{34}=0$.} In this case the $\nu_4$ ($\bar{\nu}_4$) contributes to both $\nu_e$ and $\nu_\mu$ ($\bar{\nu}_e$ and $\bar{\nu}_\mu$) and effectively the oscillation pattern is a combination of \textbf{Case I} and \textbf{Case II}; {\it i. e.}, the enhancement of active-sterile oscillation takes place for both $\nu_e$ and $\bar{\nu}_\mu$ atmospheric fluxes. In this case the deficit of atmospheric flux is more significant and sensitivity of IceCube/DeepCore to active-sterile mixing parameters is enhanced.

\begin{figure}[t!]
\begin{center}
\subfloat[$P(\bar{\nu}_\mu\to\bar{\nu}_\tau)$ for $\sin^22\theta_{24}=\sin^22\theta_{34}=0.1$]{
 \includegraphics[width=0.5\textwidth]{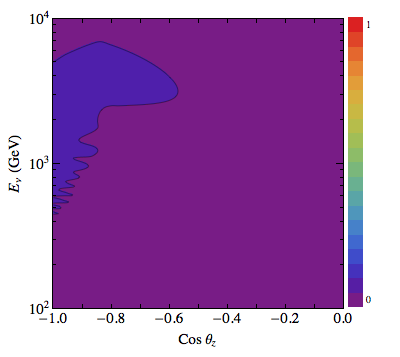}
  \label{fig:mubartotaubar,21,1,8,8}
}
\subfloat[$P(\bar{\nu}_\mu\to\bar{\nu}_\tau)$ for $\sin^22\theta_{24}=0.1$ and $\sin^22\theta_{34}=0.3$]{
 \includegraphics[width=0.5\textwidth]{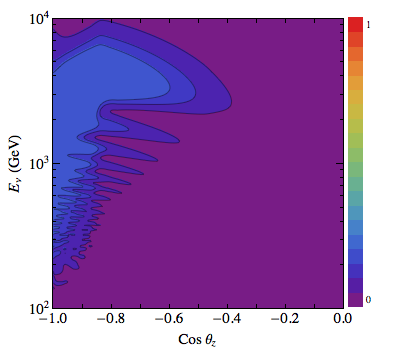}
 \label{fig:mubartotaubar,21,1,8,11}
}
\quad
\subfloat[$P(\bar{\nu}_\mu\to\bar{\nu}_\tau)$ for $\sin^22\theta_{24}=0.3$ and $\sin^22\theta_{34}=0.1$]{
 \includegraphics[width=0.5\textwidth]{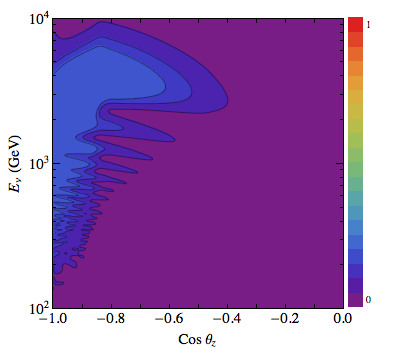}
  \label{fig:mubartotaubar,21,1,11,8}
}
\subfloat[$P(\bar{\nu}_\mu\to\bar{\nu}_\tau)$ for $\sin^22\theta_{24}=\sin^22\theta_{34}=0.3$]{
 \includegraphics[width=0.5\textwidth]{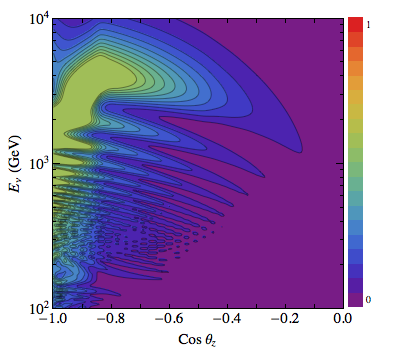}
 \label{fig:mubartotaubar,21,1,11,11}
}
\end{center}
\caption{\label{fig:probcaseVI}Oscillogram of oscillation probability $P(\bar{\nu}_\mu\to\bar{\nu}_\tau)$ for $\Delta m_{41}^2=1\,{\rm eV}^2$, $\theta_{14}=0$ and different values of $\sin^22\theta_{24}$ and $\sin^22\theta_{34}$ indicated in each sub-caption.}
\end{figure}

\paragraph{Case V: $\theta_{14}\neq0$, $\theta_{24}=0$ and $\theta_{34}\neq0$.} In this case the enhancement of active-sterile oscillation takes place for $\nu_e$ and $\bar{\nu}_\tau$ fluxes. However, as we discussed in {\textbf{Case III}, the atmospheric $\bar{\nu}_\tau$ flux is negligible in this energy range and sensitivity of IceCube/DeepCore in this case is similar to \textbf{Case I}.

\paragraph{Case VI: $\theta_{14}=0$, $\theta_{24}\neq0$ and $\theta_{34}\neq0$.} In this case the contribution of $\nu_4$ ($\bar{\nu}_4$) to both $\nu_\mu$ and $\nu_\tau$ ($\bar{\nu}_\mu$ and $\bar{\nu}_\tau$) is nonzero. Thus, in this case we expect resonant $\bar{\nu}_\mu\to\bar{\nu}_s$ and $\bar{\nu}_\tau\to\bar{\nu}_s$ conversions, although the latter is absent due to the negligible atmospheric flux of $\bar{\nu}_\tau$. However, an interesting phenomenon is that, due to the nonzero value of $\theta_{34}$ in this case, $\bar{\nu}_\mu$ do not completely convert to $\bar{\nu}_s$; but partially converts to $\bar{\nu}_\tau$. Thus, we effectively observe $\bar{\nu}_\mu\to \bar{\nu}_\tau$ conversion, indirectly induced by $\bar{\nu}_s$; this phenomenon has been discussed in~\cite{Choubey:2007ji}. To illustrate this phenomenon, in Figure~\ref{fig:probcaseVI} we show the oscillogram of $P(\bar{\nu}_\mu\to\bar{\nu}_\tau)$ for various values of $\theta_{24}$ and $\theta_{34}$. As can be seen, the nonzero value of $\theta_{34}$ leads to $\bar{\nu}_\mu\to\bar{\nu}_\tau$ conversion; while in $3\nu$ framework, in the energy range we have shown, we have $P(\bar{\nu}_\mu\to\bar{\nu}_\tau)=0$. Through this effect, the energy and zenith distributions of cascades in IceCube/DeepCore are sensitive to $\theta_{34}$. We emphasize that the sensitivity to $\theta_{34}$ is through $\bar{\nu}_\mu\to\bar{\nu}_\tau$ conversion introduced in this case. Thus, to constrain $\theta_{34}$, the necessary condition is $\theta_{24}\neq0$, which is hinted by MiniBooNE and LSND anomalies.

\paragraph{Case VII: $\theta_{14}\neq0$, $\theta_{24}\neq0$ and $\theta_{34}\neq0$.} In this case all the active-sterile mixing angles are nonzero and $\nu_4$ ($\bar{\nu}_4$) contributes to all the active (anti-)neutrino flavor states. Thus, by analyzing zenith and energy distributions of atmospheric induced cascade events it is possible to constrain all the active-sterile mixing angles $\theta_{14}$, $\theta_{24}$ and $\theta_{34}$; respectively through enhanced $\nu_e\to\nu_s$, $\bar{\nu}_\mu\to\bar{\nu}_s$ and $\bar{\nu}_\mu\to\bar{\nu}_\tau$ conversions.

For the analysis of this paper we numerically calculated all oscillation probabilities $P(\nu_\alpha(\bar{\nu}_\alpha)\to\nu_\beta(\bar{\nu}_\beta))$ as a function of $\cos \theta_z$ and $E_\nu$, scanning the whole parameter space of $(\Delta m_{41}^2,\sin^2 2\theta_{14},\sin^2 2\theta_{24},\sin^2 2\theta_{34})$, for more than 26,000 sets of mixing parameter values. In the next section we calculate sensitivity of DeepCore to $3+1$ model by performing a $\chi^2$-analysis of zenith and energy distributions of atmospheric induced cascade events.

\section{\label{sec:limits}Sensitivity of DeepCore to active-sterile mixing angles $\theta_{i4}$}

With the oscillation probabilities in $3+1$ model, discussed in section~\ref{sec:matterprob}, we can calculate number of cascade events in DeepCore induced by the atmospheric flux of neutrinos. DeepCore is the inner region of IceCube which is more densely instrumented with photomultipliers of higher quantum efficiency. Higher efficiency of photomultipliers, less distance between them, clearer ice in the deep regions of detector, and using the remainder of IceCube detector as a veto for atmospheric muon background, leads to a lower energy threshold for the DeepCore part of IceCube. Detection of cascade events have been reported by IceCube collaboration~\cite{Abbasi:2011ui,Aartsen:2012uu,:2012zk}. In early DeepCore data~\cite{:2012zk} cascades have been detected with energy as low as $\sim 30$~GeV. In this section we calculate sensitivity of DeepCore to the mixing angles $\theta_{i4}$ ($i=1,2,3$), using a $\chi^2$-analysis. We present here a first exploratory analysis illustrating the physics potential of cascade events in probing active-sterile neutrino mixing. Here we account for the detector response (energy and direction reconstruction resolution, ... ) by the very coarse binning of data. However, it cannot be a substitute for a detailed analysis which will include smearing of fine binned data. In section~\ref{sec:chi2}, after discussing the rate of cascade events in DeepCore, we define the $\chi^2$ function of our analysis. In section~\ref{sec:q1424} we show sensitivity to $\theta_{14}$ and $\theta_{24}$ mixing angles. Finally, in section~\ref{sec:q34}, we show sensitivity of DeepCore to $\theta_{34}$, which is the main result of this paper.

\subsection{\label{sec:chi2}Number of cascades and $\chi^2$ function}

As we discussed in section~\ref{sec:matterprob}, cascade events originate from the following interactions: (i) NC interaction of all active flavor neutrinos; (ii) CC interaction of $\nu_e$ and $\bar{\nu}_e$; (iii) CC interaction of $\nu_\tau$ and $\bar{\nu}_\tau$. The number of cascade events from (i) is 

$$N^{{\rm NC}}_{\rm cas}=N^{{\rm NC},\nu_e~{\rm atm}}_{\rm cas}+N^{{\rm NC},\nu_\mu~{\rm atm}}_{\rm cas}~,$$ 
where
\begin{eqnarray}
N^{{\rm NC},\nu_e(\nu_\mu)~{\rm atm}}_{\rm cas} & = T\Delta \Omega \rho_{\rm ice} N_A  & \sum_{\alpha=e,\mu,\tau} \int \sigma^{\rm NC}(E_\nu) \Phi^{\rm atm}_{\nu_e(\nu_\mu)}(E_\nu,\cos\theta_z)\times \\
& & P(\nu_e(\nu_\mu)\to\nu_\alpha) V^{\rm DC}_{\rm eff} (E_\nu,\cos\theta_z) {\rm d}E_\nu {\rm d}\cos\theta_z + (\nu\to\bar{\nu})~.\nonumber 
\end{eqnarray}
In this equation $T$ is the live-time of data-taking, $\Delta \Omega=2\pi$ is the azimuthal acceptance of DeepCore, $\rho_{\rm ice}$ is the ice density, $N_A$ is the Avogadro's number and $\sigma^{\rm NC}$ is the neutral current cross section of neutrinos. For the flux of atmospheric electron (muon) neutrinos, $\Phi^{\rm atm}_{\nu_e(\nu_\mu)}(E_\nu,\cos\theta_z)$, we use Honda flux~\cite{Honda:2006qj}. $P$ represents oscillation probabilities calculated in section~\ref{sec:matterprob}. $V^{\rm DC}_{\rm eff}$ is the effective volume of DeepCore for cascade detection. Since DeepCore detector resides at the inner part of IceCube detector, to a good approximation the effective volume is independent of zenith angle of incoming neutrinos and $V^{\rm DC}_{\rm eff}(E_\nu,\cos \theta_z)\equiv V^{\rm DC}_{\rm eff}(E_\nu)$. For the value of effective volume we consider two extreme cases $V_{\rm eff}^{\rm pes}$ and $V_{\rm eff}^{\rm opt}$, denoting respectively a pessimistic and optimistic estimation of effective volume. For $V_{\rm eff}^{\rm opt}$ we use the ``online filter" effective volume from~\cite{Collaboration:2011ym} and for $V_{\rm eff}^{\rm pes}$ we use effective volume of~\cite{:2012zk}. The latter corresponds to an analysis of DeepCore data using IceCube analysis tools that have not been optimized for the analysis of low energy events. Figure~\ref{fig:Veff} shows the effective volumes used in this paper and it should be noticed that the realistic effective volume of DeepCore, after developing appropriate quality cuts, will lie between these two cases.

\begin{figure}[t!]
\begin{center}
\includegraphics[width=0.7\textwidth]{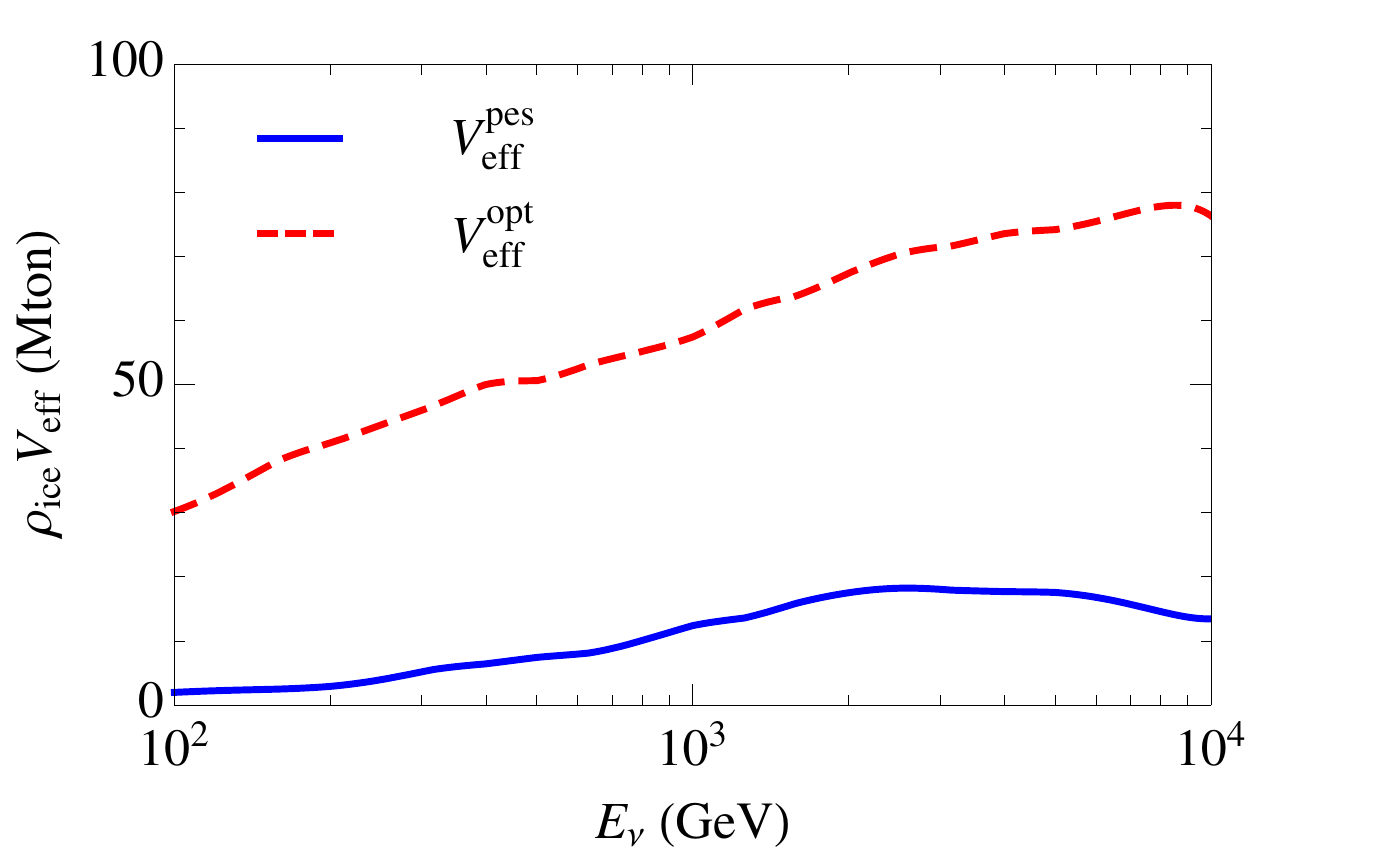}
\end{center}
\caption{\label{fig:Veff}The effective volumes used in this paper. $V_{\rm eff}^{\rm opt}$ refers to the ``online filter" effective volume of DeepCore from~\cite{Collaboration:2011ym} and $V_{\rm eff}^{\rm pes}$ refers to the effective volume from~\cite{:2012zk}.}
\end{figure}

The number of cascade events from (ii) and (iii) are $N^{{\rm CC},e}_{\rm cas}$ and $N^{{\rm CC},\tau}_{\rm cas}$, respectively, given by
\begin{eqnarray}
N^{{\rm CC},e(\tau)}_{\rm cas} & = T\Delta \Omega \rho_{\rm ice} N_A  & \sum_{\alpha=e,\mu} \int \sigma^{\rm CC}(E_\nu) \Phi^{\rm atm}_{\nu_\alpha}(E_\nu,\cos\theta_z)\times \nonumber \\
& & P(\nu_\alpha\to\nu_e(\nu_\tau)) V^{\rm DC}_{\rm eff} (E_\nu,\cos\theta_z) {\rm d}E_\nu {\rm d}\cos\theta_z + (\nu\to\bar{\nu})~.
\end{eqnarray}
Total number of cascade events at DeepCore, $N_{\rm cas}$, is the sum of numbers from (i), (ii) and (iii); {\it i. e.}, $N_{\rm cas}=N^{{\rm NC}}_{\rm cas}+N^{{\rm CC},e}_{\rm cas}+N^{{\rm CC},\tau}_{\rm cas}$. In order to calculate sensitivity of DeepCore to the parameter space of $3+1$ model, we define the following $\chi^2$ function:
\begin{eqnarray}\label{eq:chi2}
\chi^2 (\{\Delta m_{41}^2, \theta_{i4};\alpha,\beta)  = \qquad \qquad \qquad \qquad \qquad \qquad \qquad \qquad \qquad \qquad \qquad \qquad \qquad \qquad \quad & &  \\
\sum_{i,j} \frac{\left[N^{i,j}_{\rm cas}(\{\Delta m_{41}^2, \theta_{i4}\}^{\rm true}=0)-\alpha\left( 1+\beta(0.5+\cos\theta_z) \right)N^{i,j}_{\rm cas}(\{\Delta m_{41}^2, \theta_{i4}\})\right]^2}{N^{i,j}_{\rm cas}(\{\Delta m_{41}^2, \theta_{i4}\}^{\rm true} =0)}+\frac{(1-\alpha)^2}{\sigma_\alpha^2}+\frac{\beta^2}{\sigma_\beta^2} & & \nonumber
\end{eqnarray}
where $\{\Delta m_{41}^2, \theta_{i4}\}\equiv(\Delta m_{41}^2,\theta_{14},\theta_{24},\theta_{34})$ denotes the active-sterile mixing parameters, $\alpha$ and $\beta$ are the parameters which take into account respectively normalization and zenith dependent uncertainties of atmospheric neutrino flux, with $\sigma_\alpha=0.24$ and $\sigma_\beta=0.04$. In the $\chi^2$ defined in eq.~(\ref{eq:chi2}) we assumed vanishing true values for active-sterile mixing parameters and thus this function provides the sensitivity of IceCube/DeepCore to nonzero active-sterile mixing parameters. The summation indices $i$ and $j$ run over neutrino energy and zenith bins, respectively; such that $N^{i,j}_{\rm cas}$ shows the number of cascade events with initial neutrino energy in the $i$-th bin of energy and $\cos\theta_{z}$ in the $j$-th bin of zenith distribution. DeepCore detector acts like a calorimeter for cascade events with great precision in the measurement of released energy in cascades. In the analysis of IC-22 cascades, IceCube collaboration reported $\Delta(\log_{10}(E_\nu/{\rm GeV}))=0.18$ for the resolution of energy reconstruction~\cite{Abbasi:2011ui}. In the analysis of this paper, as a realistic energy resolution for cascade events at DeepCore, we assume $\Delta(\log_{10}(E_\nu/{\rm GeV}))=0.1$ (see also~\cite{icrc2009,thesis}). Thus, totally we have 20 bins of energy from 100~GeV to 10~TeV, with the width $\log_{10}(E_\nu/{\rm GeV})=0.1$. In spite of the good resolution in energy reconstruction, direction reconstruction of cascades in IceCube/DeepCore is poor. However, by performing improved algorithms for the direction reconstruction, it is possible to reach direction resolution as low as $25^\circ$~\cite{icrc2009,thesis}. In this paper we calculate sensitivity of DeepCore assuming different values for direction resolution: assuming $45^\circ$ and $30^\circ$ for direction resolution, we construct ``two bins" ($[-1,-0.7] , [-0.7,0]$) and ``three bins" ($[-1,-0.9] , [-0.9,-0.5] , [-0.5,0]$) of $\cos\theta_z$, respectively. Also, to find out the reward of achieving better resolution in direction reconstruction, we calculate sensitivity of DeepCore to $\theta_{i4}$ assuming 10 bins of $\cos\theta_z$ (with width 0.1). However, we emphasize that the sensitivity with 10 bins is quite optimistic and achieving resolution $0.1$ in $\cos\theta_z$ is quite challenging.

\subsection{\label{sec:q1424}Constraining $\theta_{14}$ and $\theta_{24}$}

As we discussed in \textbf{Case I}, \textbf{Case II} and \textbf{Case IV} in section~\ref{sec:matterprob}, nonzero values of $\theta_{14}$ and $\theta_{24}$ lead to distortions in $\nu_e\to\nu_e$ and $\bar{\nu}_\mu\to\bar{\nu}_\mu$ oscillation probabilities, respectively, which makes it possible to constrain these parameters through the detection of cascade events in DeepCore. In the following we show sensitivity of DeepCore to these mixing parameters.

\begin{figure}[t!]
\begin{center}
\includegraphics[width=0.6\textwidth]{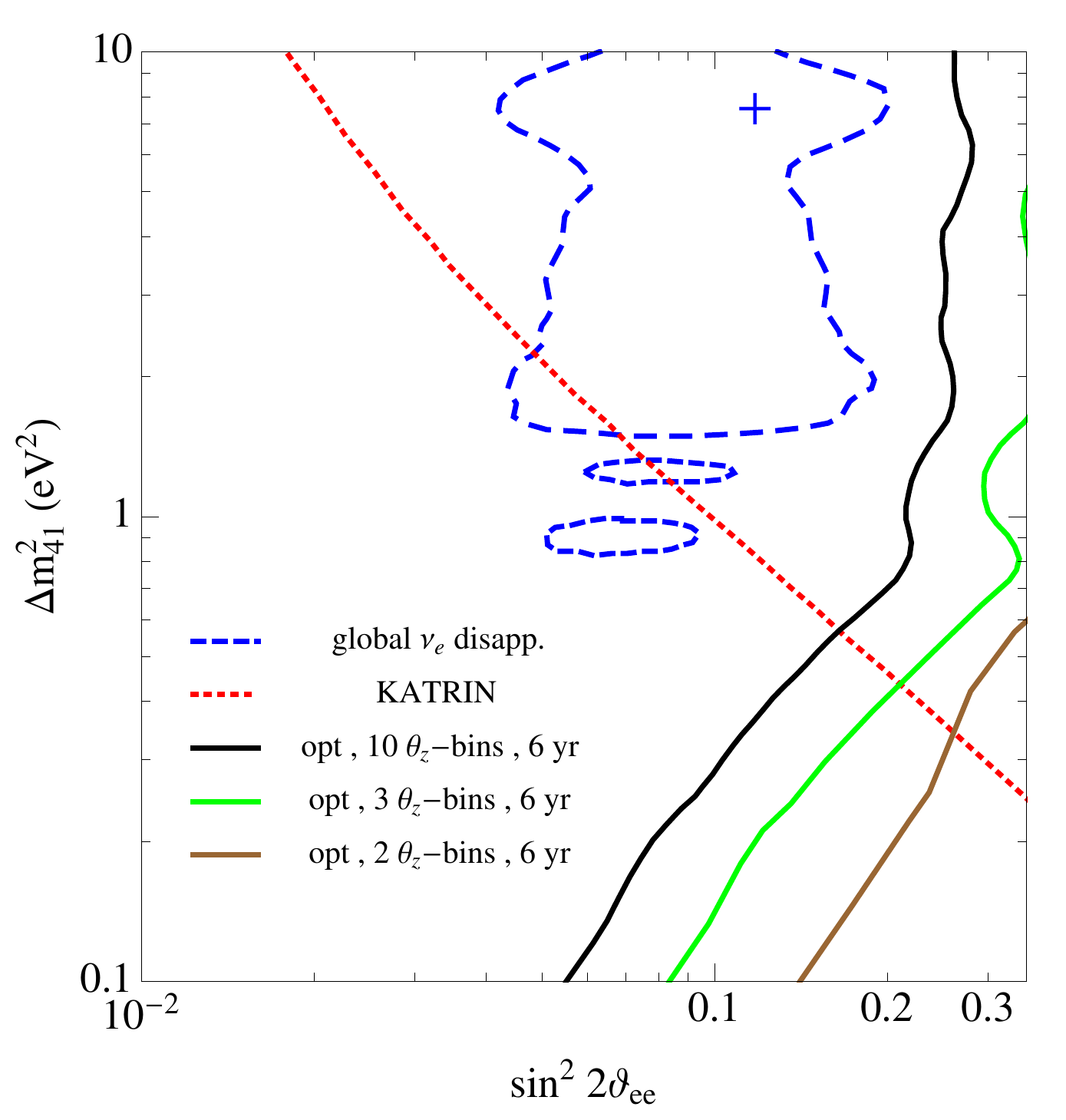}
\end{center}
\caption{\label{fig:q14}Sensitivity of DeepCore to $\vartheta_{ee}$ in the plane $(\sin^2 2\vartheta_{ee},\Delta m_{41}^2)$. The solid curves (black, green and brown, respectively for 10, 3 and 2 bins of $\cos\theta_z$) show sensitivity after 6 years of data-taking with effective volume $V_{\rm eff}^{\rm opt}$ corresponding to red dashed curve in Figure~\ref{fig:Veff}. Blue dashed curve shows the allowed region from global analysis of reactor, Gallium, $\nu_e{\rm C}$ scattering and solar data, taken from~\cite{Giunti:2012tn}. The red dotted curve shows sensitivity of KATRIN experiments after three years of data-taking, from~\cite{Esmaili:2012vg}. All the curves are at 90\% C.L.. }
\end{figure}

\begin{figure}[h!]
\begin{center}
\includegraphics[width=0.6\textwidth]{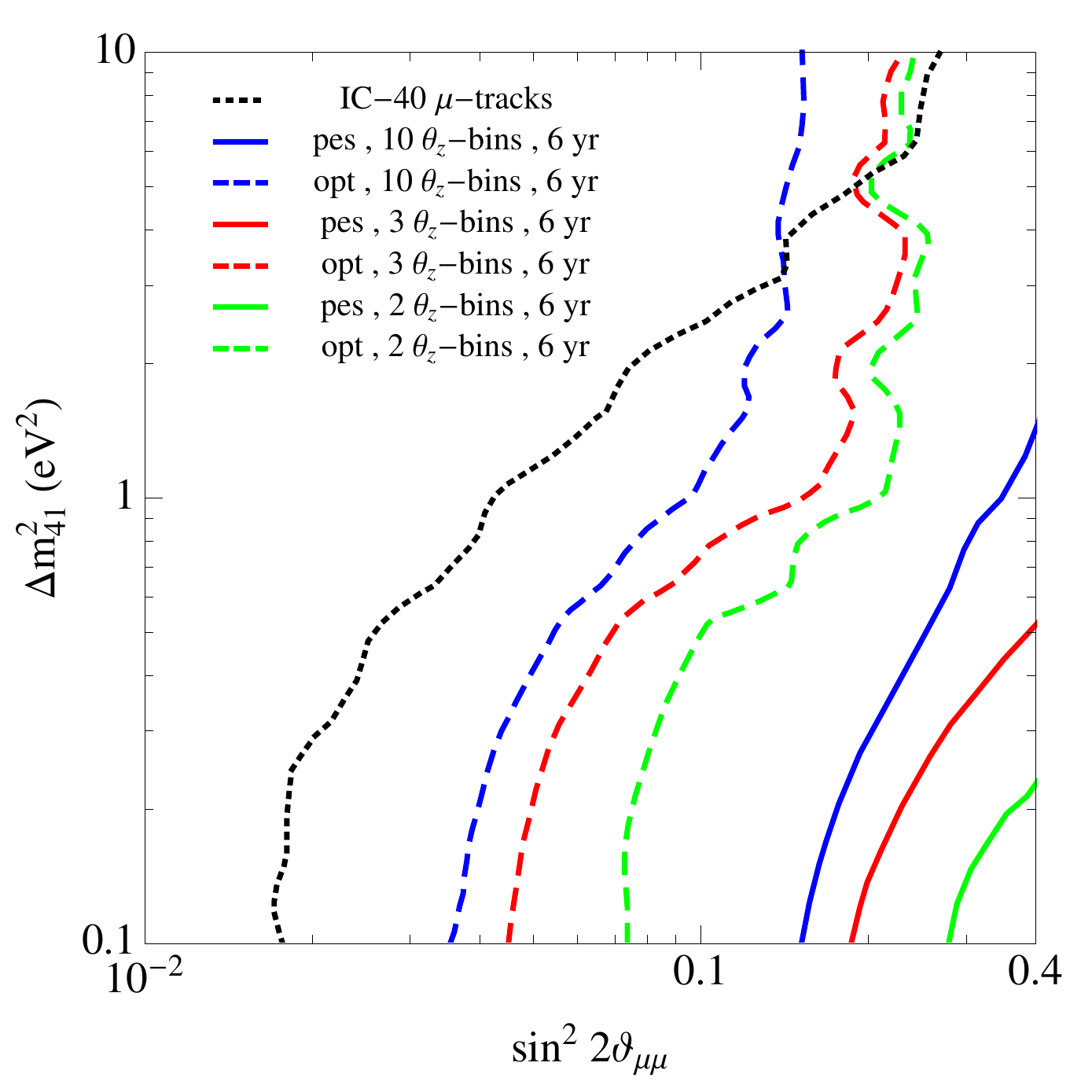}
\end{center}
\caption{\label{fig:qmumu}Sensitivity of DeepCore to $\vartheta_{\mu\mu}$ in the plane $(\sin^2 2\vartheta_{\mu\mu},\Delta m_{41}^2)$. The solid and dashed curves correspond to sensitivity of DeepCore with effective volume $V_{\rm eff}^{\rm pes}$ and $V_{\rm eff}^{\rm opt}$, respectively. The blue, red and green color correspond to sensitivity after six years of data-taking with 10, 3 and 2 bins of $\cos\theta_z$ respectively. Black dotted curve shows the best current upper limit from atmospheric neutrino data of IC-40~\cite{Esmaili:2012nz}. All the curves are at 90\% C.L..  }
\end{figure}

\paragraph{Sensitivity to $\theta_{14}$:} As we discussed in section~\ref{sec:matterprob}, sensitivity to $\theta_{14}$ originates from enhancement of $\nu_e\to\nu_s$ oscillation for nonzero values of $\theta_{14}$ (see Figure~\ref{fig:probcaseI}). However, since the flux of atmospheric $\nu_e$ is small in the high energy range that we are considering ($\Phi^{\rm atm}_{\nu_\mu+\bar{\nu}_\mu}/\Phi^{\rm atm}_{\nu_e+\bar{\nu}_e}\gtrsim 30$ for $E_\nu\gtrsim100$~GeV), the low statistics leads to a poor sensitivity to $\theta_{14}\equiv\vartheta_{ee}$. Figure~\ref{fig:q14} shows sensitivity of DeepCore in $(\sin^2 2\vartheta_{ee},\Delta m_{41}^2)$ plane, obtained after marginalizing $\chi^2$ function in eq.~(\ref{eq:chi2}) with respect to $\theta_{24}$ and $\theta_{34}$. The black, green and brown solid lines correspond to 90\% C.L. sensitivity assuming 10, 3 and 2 bins of $\cos\theta_z$, respectively. For all the three solid curves we used $V_{\rm eff}^{\rm opt}$ (the red dashed curve in Figure~\ref{fig:Veff}) and six years of data-taking. Blue dashed curve shows the 90\% C.L. allowed region in $(\sin^2 2\vartheta_{ee},\Delta m_{41}^2)$ from analysis of reactor, Gallium, $\nu_e{\rm C}$ scattering and solar data, taken from~\cite{Giunti:2012tn}. Red dotted curve of Figure~\ref{fig:q14} shows the 90\% C.L. sensitivity of KATRIN experiment (measuring the spectrum of electrons from tritium beta decay) after 3 years of data-taking, from~\cite{Esmaili:2012vg}. As can be seen, even with the optimistic assumption for effective volume of DeepCore and very low uncertainty in direction reconstruction of cascades, sensitivity of DeepCore is not enough to cover the favored region by short baseline experiments. However, with the realistic assumption of two bins for $\cos\theta_z$ (brown solid curve) it is possible to exclude parameter space near $\Delta m_{41}^2\sim0.1-0.3~{\rm eV}^2$ better than KATRIN. With the pessimistic assumption for effective volume of DeepCore (blue solid curve in Figure~\ref{fig:Veff}), we do not have sensitivity to $\sin^2 2\vartheta_{ee}$.

\paragraph{Sensitivity to $\theta_{24}$:} Sensitivity of DeepCore to $\theta_{24}$ originates from enhancement of $\bar{\nu}_\mu\to\bar{\nu}_s$ oscillation for $\theta_{24}\neq0$, which leads to a decrease in the number and distortion in energy and zenith distributions of cascade events. In the calculation of sensitivity to $\theta_{24}$ we assume $\theta_{14}=0$, which implies: $\theta_{24}\equiv\vartheta_{\mu\mu}$, and marginalize $\chi^2$ function in eq.~(\ref{eq:chi2}) with respect to $\theta_{34}$. Figure~\ref{fig:qmumu} shows the sensitivity in $(\sin^2 2\vartheta_{\mu\mu},\Delta m_{41}^2)$ plane, where solid and dashed lines correspond respectively to $V_{\rm eff}^{\rm pes}$ and $V_{\rm eff}^{\rm opt}$ for effective volume; and blue, red and green colors correspond to 10, 3 and 2 bins of $\cos\theta_z$, respectively. Black dotted curve in Figure~\ref{fig:qmumu} shows the strongest current upper limit from the analysis of $\mu$-tracks induced by high energy atmospheric neutrinos at IceCube-40~\cite{Esmaili:2012nz}. However, as we expect, due to the high statistics of $\mu$-track events, sensitivity calculated from cascade events at DeepCore, even with optimistic assumptions, is less than the IceCube-40 upper limit. Also, it should be noticed that the upper limit in~\cite{Esmaili:2012nz}; {\it i.~e.}, black dotted curve in Figure~\ref{fig:qmumu}, is calculated by fitting zenith distribution of energy integrated $\mu$-track data from~\cite{Abbasi:2010ie} and definitely, analyzing the data with energy binning can improve the limit significantly.

\subsection{\label{sec:q34}Constraining $\theta_{34}$}

As we discussed in \textbf{Case VI} in section~\ref{sec:matterprob}, nonzero values of both $\theta_{24}$ and $\theta_{34}$ lead to $P(\bar{\nu}_\mu\to\bar{\nu}_\tau)\neq0$, which consequently distorts zenith and energy distributions of cascade events in DeepCore. Thus, in principle, by looking at zenith and energy distributions of cascade events in DeepCore, it is possible to constrain $\theta_{34}$ when $\theta_{24}\neq0$. It should be noticed that zenith and energy distributions of $\mu$-track events in IceCube would hint or constrain $\theta_{24}$~\cite{Razzaque:2012tp,Razzaque:2011ab,Barger:2011rc,Halzen:2011yq,Esmaili:2012nz}; thus, by the combined analysis of $\mu$-track and cascade events, it is possible to establish first the hint or limit on the value of $\theta_{24}$ and then constrain $\theta_{34}$. In order to show the effect of nonzero $\theta_{34}$, we plot $\chi^2$ function in eq.~(\ref{eq:chi2}) with respect to $\sin^22\theta_{34}$ in Figures~\ref{fig:chi2,q34,2bin} and \ref{fig:chi2,q34,3bin}, respectively for two and three bins of $\cos\theta_z$. In this figures $\sin^22\theta_{24}$ and $\Delta m_{41}^2$ are fixed to values showed in legends; and we set $\theta_{14}=0$. The true values of all the active-sterile mixing parameters set to zero. Comparing curves with the same color from Figures~\ref{fig:chi2,q34,2bin} and \ref{fig:chi2,q34,3bin} shows that, as we expect, for three bins of $\cos\theta_z$ sensitivity of DeepCore increases with respect to two bins. Comparison between blue and green curves in each plot shows that by increasing the value of $\sin^22\theta_{24}$, sensitivity of DeepCore to $\sin^22\theta_{34}$ increases. Also, position of the red curve in each plot relative to green and blue curves shows that by increasing $\Delta m_{41}^2$ sensitivity of DeepCore decreases; which is a result of lower statistics in higher energies.

\begin{figure}[t!]
\begin{center}
\subfloat[two bins of $\cos\theta_z$]{
 \includegraphics[width=0.5\textwidth]{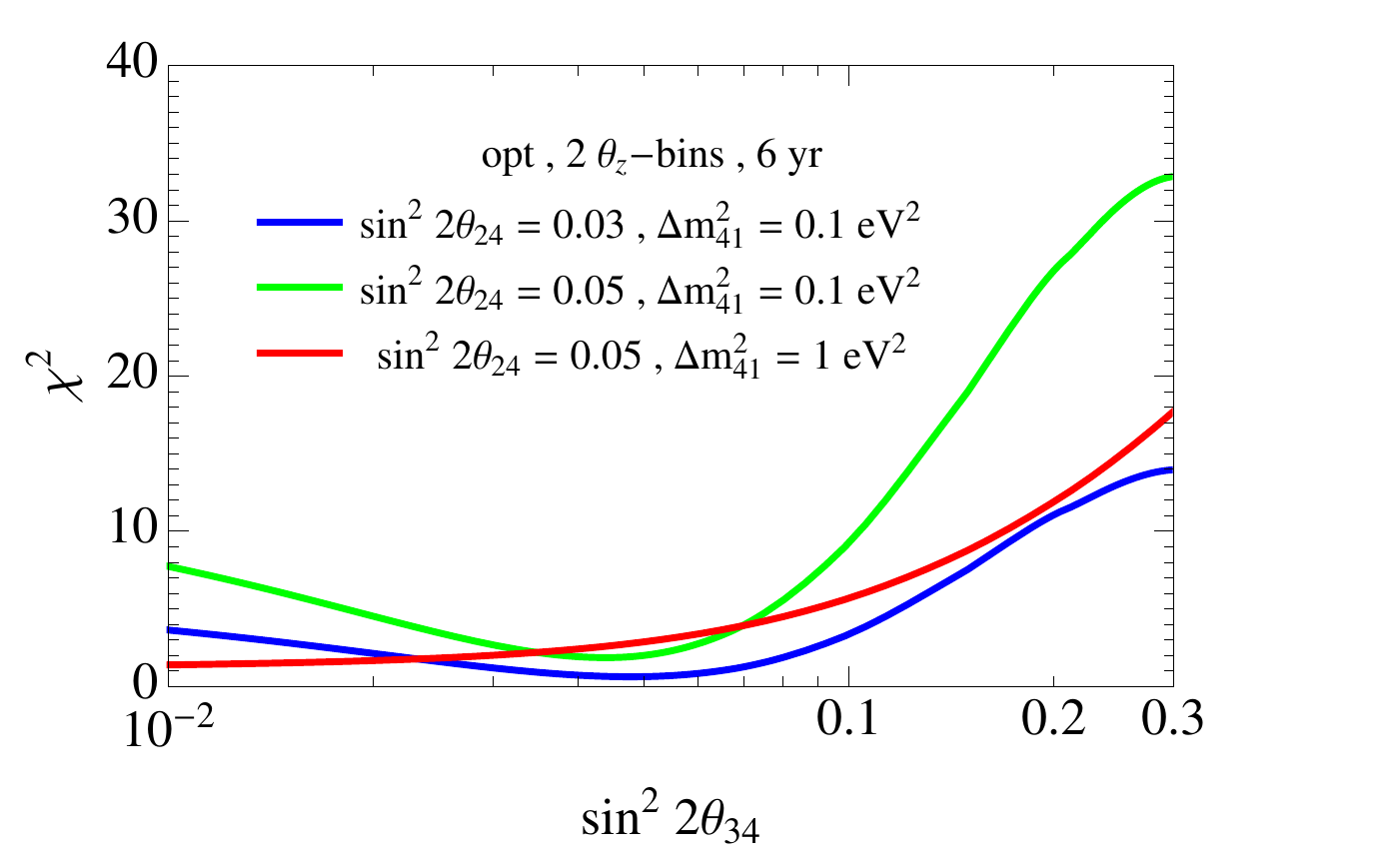}
  \label{fig:chi2,q34,2bin}
}
\subfloat[three bins of $\cos\theta_z$]{
 \includegraphics[width=0.5\textwidth]{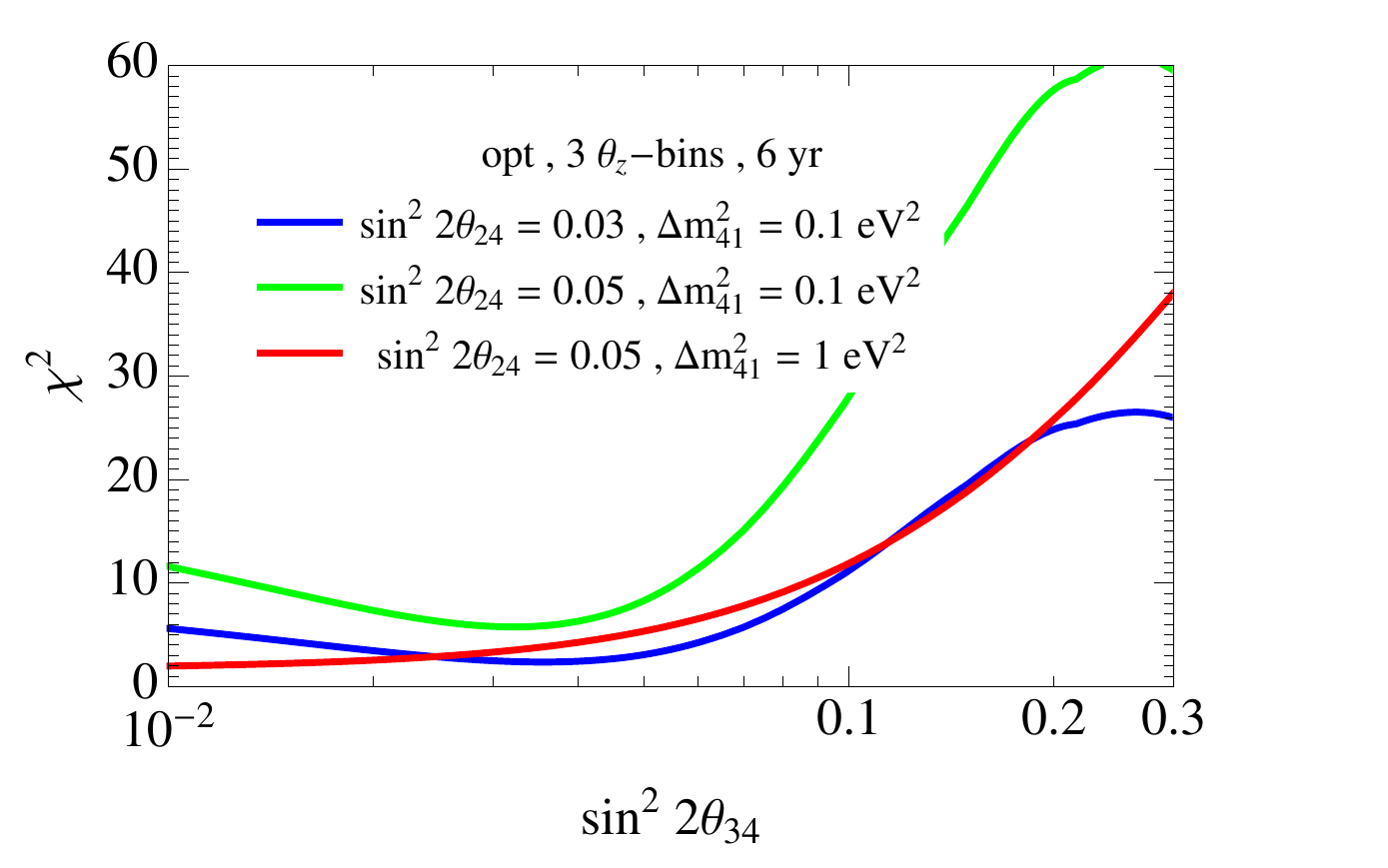}
 \label{fig:chi2,q34,3bin}
}
\end{center}
\caption{\label{fig:chi2,q34}The value of $\chi^2$ in eq.~(\ref{eq:chi2}) as a function of $\sin^2 2\theta_{34}$. The left and right plots correspond respectively to two and three bins of $\cos\theta_z$, and in both plots $V_{\rm eff}^{\rm opt}$ used for effective volume. In these plots we assumed $\theta_{14}=0$, and values of $\sin^22\theta_{24}$ and $\Delta m_{41}^2$ are shown in legends. The true values of all mixing parameters set to zero.}
\end{figure}

Two features of the curves in Figure~\ref{fig:chi2,q34} require more discussion. First, is the behavior of curves for $\theta_{34}\to0$, which as we expect, give a nonzero value for $\chi^2$ which depends on the values of $\sin^2 2\theta_{24}$ and $\Delta m_{41}^2$. This behavior is due to the fact that $\chi^2 \to 0$ when all the mixing parameters $\{\Delta m_{41}^2,\theta_{i4}\}\to 0$, while in Figure~\ref{fig:chi2,q34} we fixed $\theta_{24}$ and $\Delta m^2_{41}$ to nonzero values. The minimum of $\chi^2$ at a non-vanishing $\theta_{34}$ is the second feature of curves in Figure~\ref{fig:chi2,q34}, that we elaborate more on it here. As can be seen, this feature is obvious for low $\Delta m_{41}^2$ and by increasing $\Delta m_{41}^2$, due to low statistics, it disappears. To grasp the physics behind these minima, in Figure~\ref{fig:ratio} we show energy distribution of cascades in our ``two bins" analysis. The vertical axis in Figure~\ref{fig:ratio} is the ratio of number of cascade events in $3+1$ model, with mixing parameters shown in legends, to the number of cascades in $3\nu$ framework. Obviously, for the curves closer to a unity ratio the value of $\chi^2$ is smaller. In these plots we assumed $\Delta m_{41}^2=0.1~{\rm eV}^2$ and $\sin^22\theta_{24}=0.1$. In the first bin where $\cos\theta_z\in[-0.7,0]$ distortion in the energy distribution of events in very small. But, however, for events with $\cos\theta_z\in[-1,-0.7]$, the change in distribution is quite obvious. Important feature of Figure~\ref{fig:ratio,bin2} is the pattern of change in the curves for various values of $\sin^22\theta_{34}$. As can be seen, the red curve corresponding to $(\sin^22\theta_{24}=0.1,\sin^22\theta_{34}=0.02)$ is closer to a straight line than the black curve for $(\sin^22\theta_{24}=0.1,\sin^22\theta_{34}=0)$; which leads to a lower $\chi^2$ value for red curve with respect to black curve. The reason for this behavior is that for nonzero $\theta_{34}$ some part of initial $\bar{\nu}_\mu$ atmospheric flux converts to $\bar{\nu}_\tau$ (see Figure~\ref{fig:probcaseVI}), instead of converting to $\bar{\nu}_s$ in the case $\theta_{34}=0$, which leads to an increase in number of cascades. Thus the $\chi^2$ function has a minimum for set of $(\theta_{24},\theta_{34})$ values that the energy distribution of cascade events mimics the energy distribution in $3\nu$ framework.

One more interesting characteristic of energy distribution of cascade events in 3+1 model is the following: by increasing the value of $\sin^22\theta_{34}$ in Figure~\ref{fig:ratio,bin2}, and therefore converting almost all $\bar{\nu}_\mu$ to $\bar{\nu}_\tau$, we can see the excess in cascade events instead of deficit (see the brown curve in Figure~\ref{fig:ratio,bin2} for $\sin^22\theta_{34}=0.3$). The excess stems from higher detection rate of $\bar{\nu}_\tau$ (detectable by both CC and NC interactions) than $\bar{\nu}_\mu$ which can be detected as cascade event just by NC interaction. We would like to emphasize that this excess in the cascade events is a unique signature for $\theta_{34}\neq0$, which can be easily recognized in the experiment.

\begin{figure}[t!]
\begin{center}
\subfloat[first bin : ${\cos\theta_z  \in[-0.7,0]}$]{
 \includegraphics[width=0.5\textwidth]{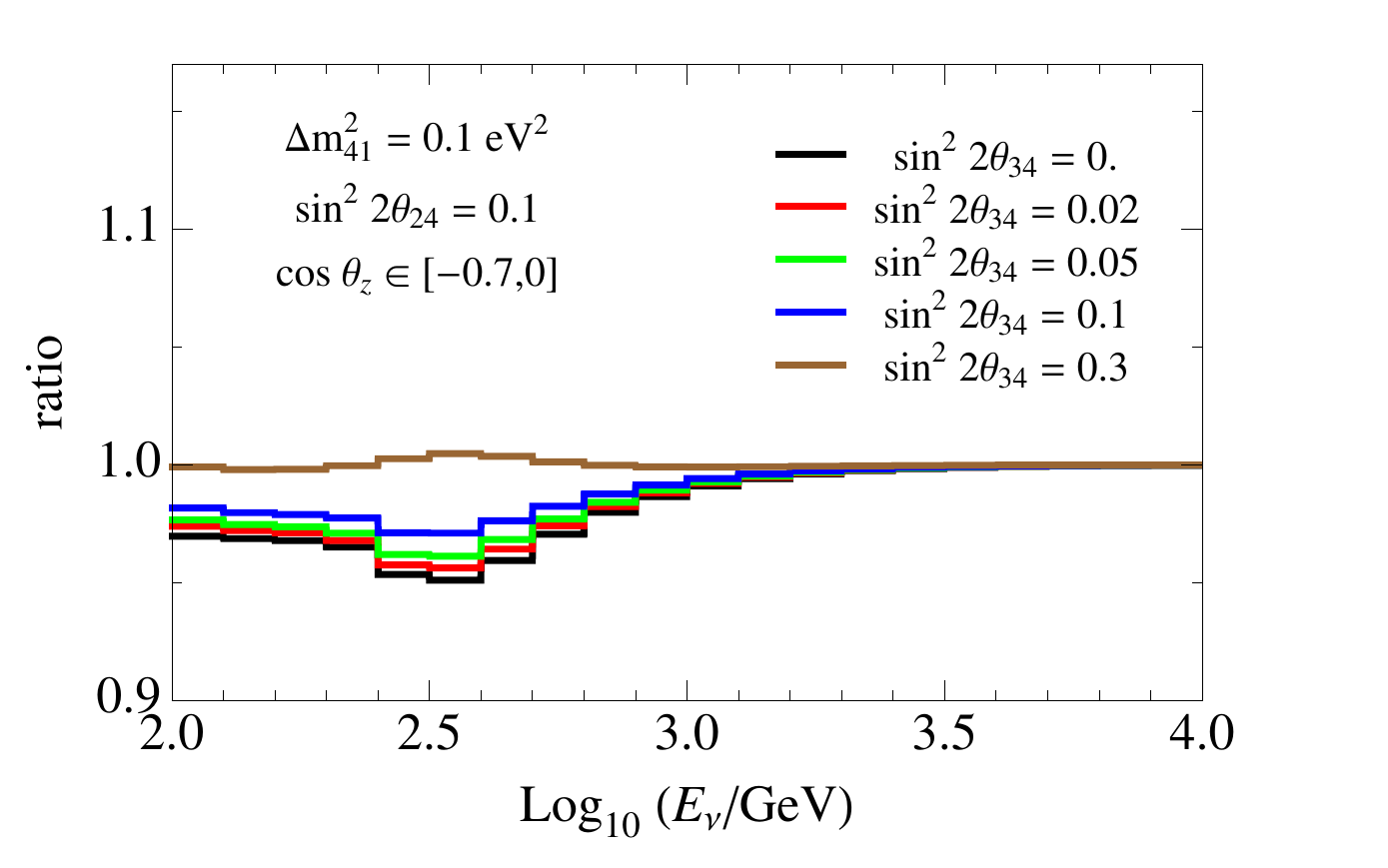}
  \label{fig:ratio,bin1}
}
\subfloat[second bin : ${\cos\theta_z  \in[-1,-0.7]}$]{
 \includegraphics[width=0.5\textwidth]{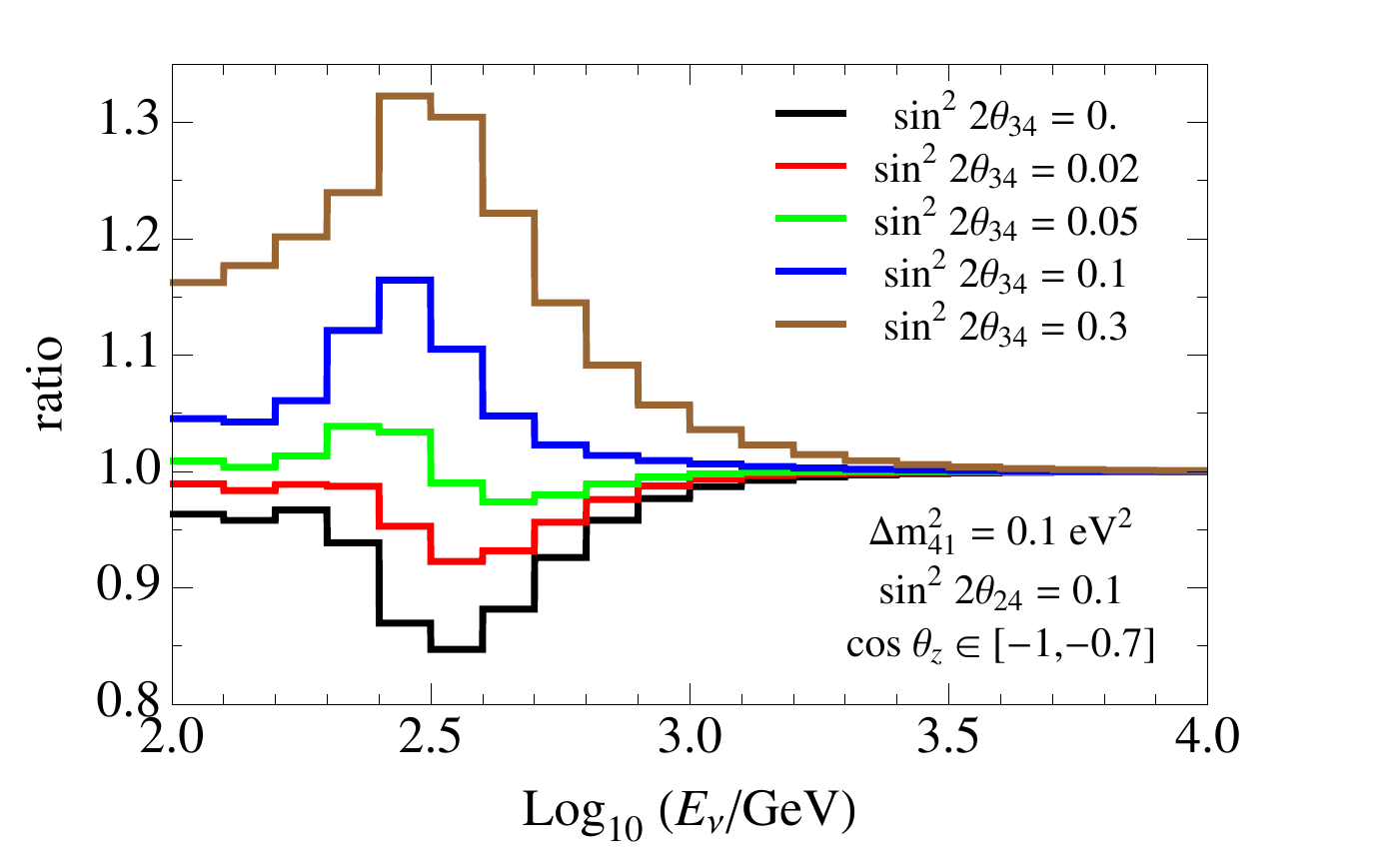}
 \label{fig:ratio,bin2}
}
\end{center}
\caption{\label{fig:ratio}Energy distribution of cascade events in the first (a) and second (b) bin of $\cos\theta_z$. The vertical axis in each plot is ratio $N_{\rm cas}(\Delta m^2_{41},\sin^2 2\theta_{24},\sin^2 2\theta_{34})/N_{\rm cas}(\Delta m^2_{41}=~0,\sin^2 2\theta_{24}=0,\sin^2 2\theta_{34}=0)$, where $N_{\rm cas}$ is the total number of cascade events and values of mixing parameters are shown in legends.}
\end{figure}

The correlation between two mixing angles $\theta_{24}$ and $\theta_{34}$ can be seen also from Figure~\ref{fig:q2434}, which shows the sensitivity of DeepCore in $(\sin^22\theta_{34},\sin^22\theta_{24})$ plane for $\Delta m_{41}^2=0.1~{\rm eV}^2$ (Figure~\ref{fig:q2434,dmspoint1}) and $\Delta m_{41}^2=1~{\rm eV}^2$ (Figure~\ref{fig:q2434,dms1}), at 90\% C.L.~. The exclusion curves in Figure~\ref{fig:q2434} obtained by assuming $\{ \Delta m_{41}^2,\theta_{i4}\}^{\rm true}=0$. The bump-shape behavior of exclusion curves for $\Delta m_{41}^2=0.1~{\rm eV}^2$ is a result of the correlation we mentioned. Also, from the curves in Figure~\ref{fig:q2434} we see that sensitivity of DeepCore to $\theta_{34}$ depends on the value of $\theta_{24}$, as we already discussed.

\begin{figure}[t!]
\begin{center}
\subfloat[$\Delta m_{41}^2=0.1~{\rm eV}^2$]{
 \includegraphics[width=0.5\textwidth]{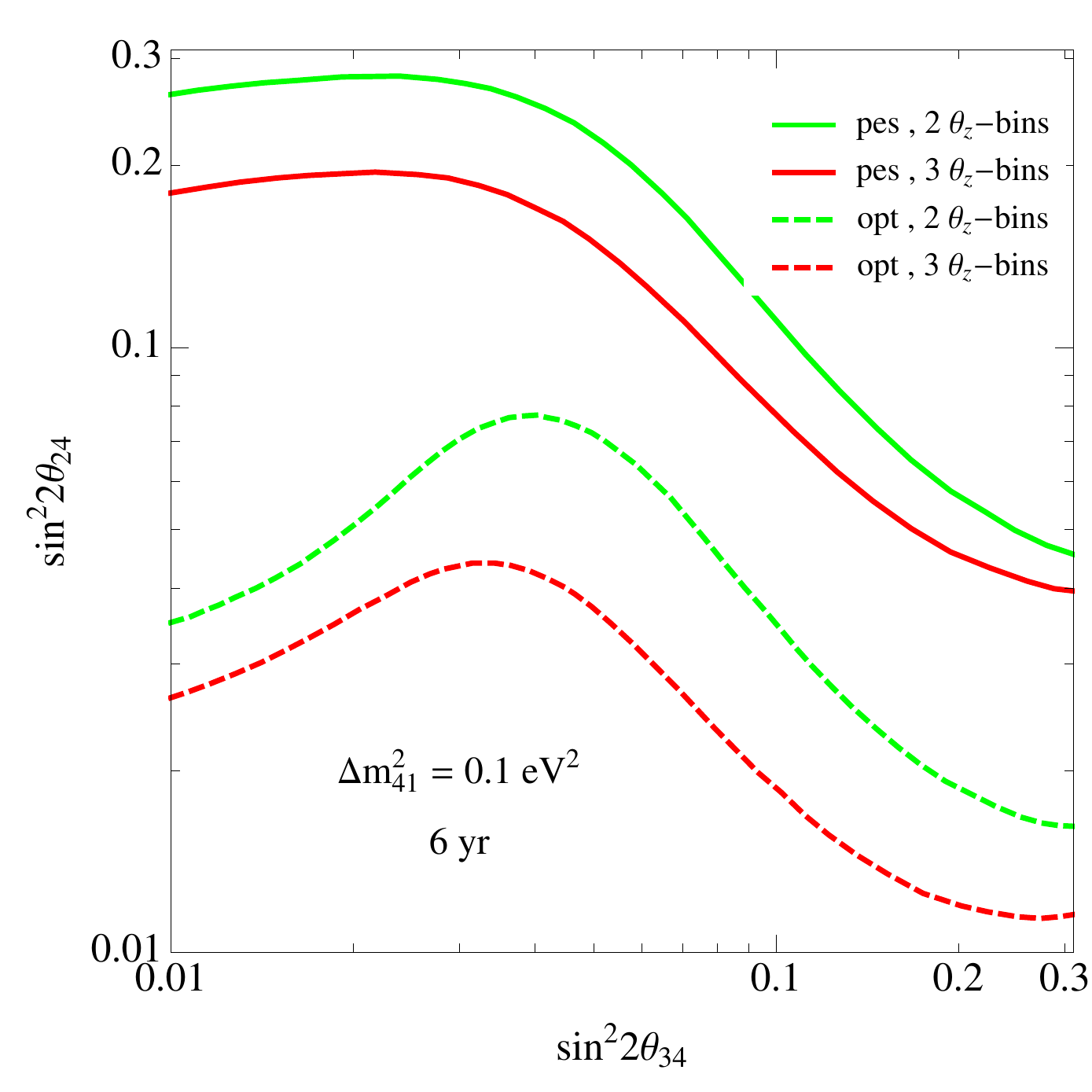}
  \label{fig:q2434,dmspoint1}
}
\subfloat[$\Delta m_{41}^2=1~{\rm eV}^2$]{
 \includegraphics[width=0.5\textwidth]{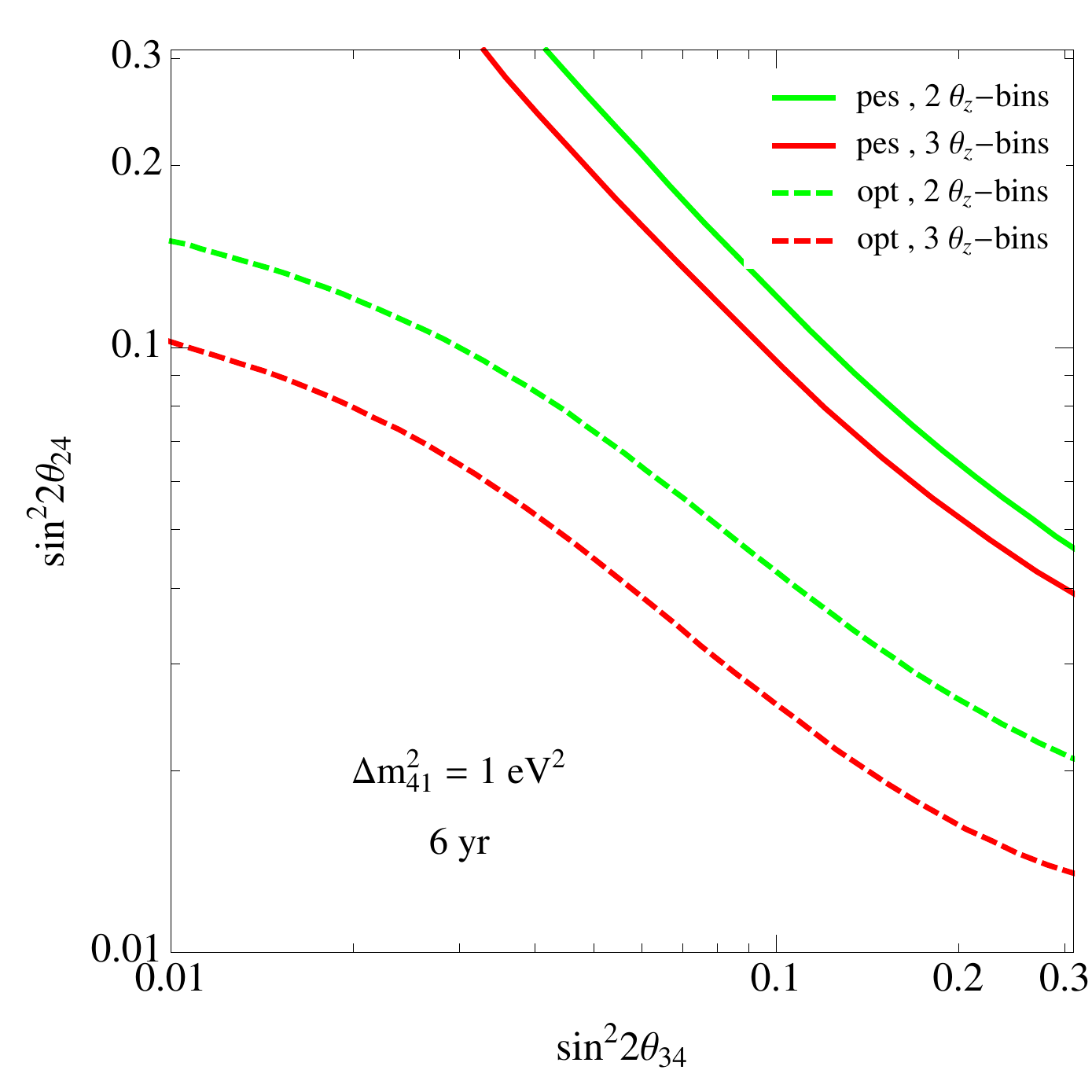}
 \label{fig:q2434,dms1}
}
\end{center}
\caption{\label{fig:q2434}Sensitivity of DeepCore in the $(\sin^22\theta_{34},\sin^22\theta_{24})$ plane for $\Delta m_{41}^2=0.1~{\rm eV}^2$ (Figure~\ref{fig:q2434,dmspoint1}) and $\Delta m_{41}^2=1~{\rm eV}^2$ (Figure~\ref{fig:q2434,dms1}). All the curves are at 90\% C.L.. In these plots we assumed $\{ \Delta m_{41}^2,\theta_{i4}\}^{\rm true}=0$.}
\end{figure}

\begin{figure}[h!]
\begin{center}
\subfloat[$\sin^22\theta_{24}^{\rm fix}=0.05 \; , \; \sin^22\theta_{34}^{\rm true}=0$]{
 \includegraphics[width=0.5\textwidth]{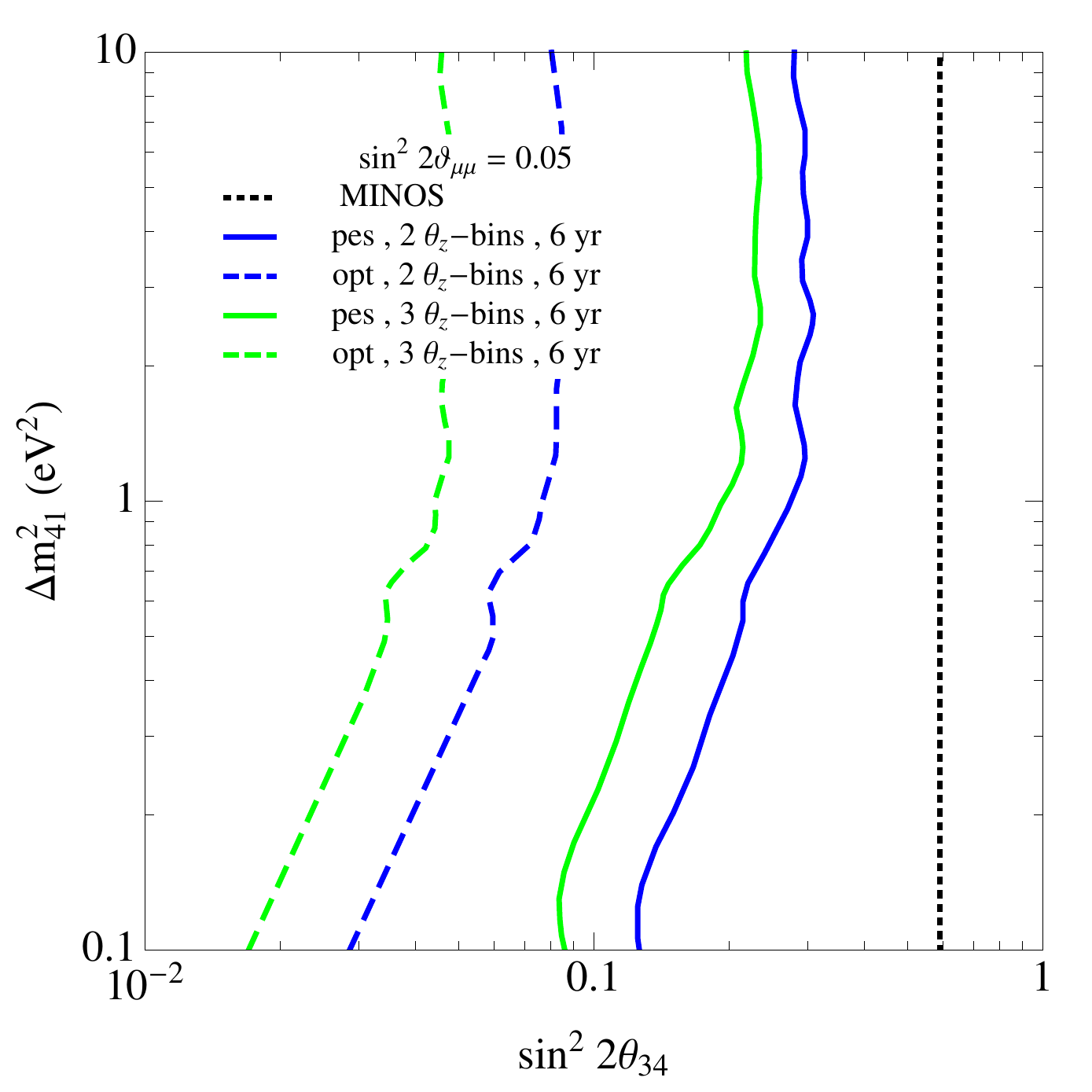}
  \label{fig:q34,q24point05}
}
\subfloat[$\sin^22\theta_{24}^{\rm fix}=0.1 \; ,\; \sin^22\theta_{34}^{\rm true}=0$]{
 \includegraphics[width=0.5\textwidth]{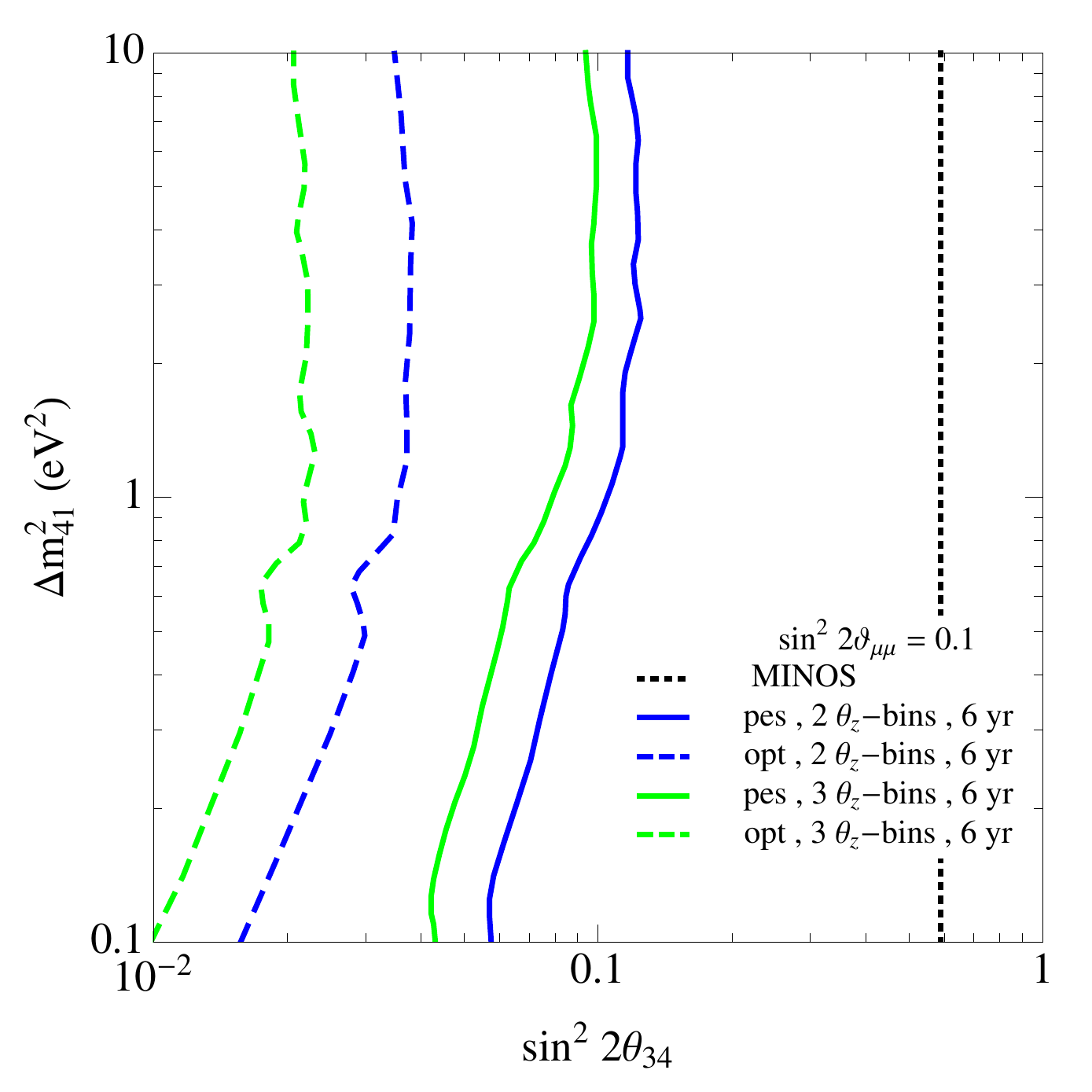}
 \label{fig:q34,q24point1}
}
\end{center}
\caption{\label{fig:q34}The sensitivity of DeepCore to $\theta_{34}$ in the plane $(\sin^2 2\theta_{34},\Delta m_{41}^2)$, obtained from $\widetilde{\chi}^2$ in eq.~(\ref{eq:tildechi2}). The solid and dashed curves correspond to sensitivity of DeepCore with effective volume $V_{\rm eff}^{\rm pes}$ and $V_{\rm eff}^{\rm opt}$, respectively. The blue and green colors correspond to sensitivity after six years of data-taking with 2 and 3 bins of $\cos\theta_z$, respectively. The black dotted vertical line shows the current upper limit from MINOS experiment~\cite{Adamson:2011ku,Adamson:2010wi}. In the left plot we assumed $\sin^22\theta_{24}^{\rm fix}=0.05$ and in the right plot $\sin^22\theta_{24}^{\rm fix}=0.1$. For all the curves $\sin^22\theta_{34}^{\rm true}=0$. All the curves are at 90\% C.L..}
\end{figure}

The non-vanishing $\chi^2$ at $\theta_{34}=0$ makes derivation of exclusion curve in $(\sin^2 2\theta_{34},\Delta m_{41}^2)$ plane ambiguous. One way of dealing with this ambiguity is to define 
$$\Delta \chi^2 \equiv \chi^2_{\theta_{14}} (\Delta m_{41}^2,\theta_{24}={\rm fixed},\theta_{34}) - \chi^2_{\theta_{14}} (\Delta m_{41}^2,\theta_{24}={\rm fixed},\theta_{34}=0)~,$$ 
where by $\chi^2_{\theta_{14}}$ we mean $\chi^2$ marginalized with respect to $\theta_{14}$, and then derive the exclusion curve from this function. However, although $\Delta\chi^2$ at $\theta_{34}=0$ is well-behaved, the minimum of $\chi^2$ in Figure~\ref{fig:chi2,q34} for nonzero values of $\sin^2 2\theta_{34}$ leads to negative values of $\Delta \chi^2$ and again makes the interpretation of $\Delta \chi^2$ ambiguous. Thus, here we define a new chi-squared function to single out sensitivity to $\theta_{34}$. In the definition of this function, $\widetilde{\chi}^2$, we assume a pre-determined value for $\theta_{24}$ from other experiments, including the IceCube $\mu$-track events which is sensitive to this parameter. Also, we marginalize with respect to $\theta_{14}$. The $\widetilde{\chi}^2$ function is defined by

\begin{eqnarray}\label{eq:tildechi2}
\widetilde{\chi}^2 (\Delta m_{41}^2,\theta_{24}={\rm fixed},\theta_{34}; \alpha,\beta)  = \frac{(1-\alpha)^2}{\sigma_\alpha^2}+\frac{\beta^2}{\sigma_\beta^2}+\qquad \qquad \qquad \qquad \qquad \qquad \qquad \qquad\qquad & &  \\
\sum_{i,j} \frac{\left[N^{i,j}_{\rm cas}(\Delta m_{41}^2, \theta_{24}^{\rm true}={\rm fixed}, \theta_{34}^{\rm true}=0)-\alpha\left( 1+\beta(0.5+\cos\theta_z) \right)N^{i,j}_{\rm cas}(\Delta m_{41}^2, \theta_{24}={\rm fixed},\theta_{34})\right]^2}
{N^{i,j}_{\rm cas}(\Delta m_{41}^2, \theta_{24}^{\rm true}={\rm fixed},\theta_{34}^{\rm true}=0)}  & & 
\nonumber
\end{eqnarray}
where the parameters $\alpha$ and $\beta$, and their corresponding uncertainties, are the same as eq.~(\ref{eq:chi2}). After marginalizing $\widetilde{\chi}^2$ with respect to $\alpha$ and $\beta$, sensitivity in the $(\sin^22\theta_{34},\Delta m_{41}^2)$ plane can be derived. Figure~\ref{fig:q34} shows the result of our analysis, for fixed $\sin^22\theta_{24}^{\rm fix}=0.05$ (Figure~\ref{fig:q34,q24point05}) and $\sin^22\theta_{24}^{\rm fix}=0.1$ (Figure~\ref{fig:q34,q24point1}), at 90\% C.L.. Dotted vertical black line shows the current upper limit $\sin^2 2\theta_{34}<0.59$ from MINOS experiment~\cite{Adamson:2011ku,Adamson:2010wi}, which we discussed in section~\ref{sec:vacuumprob}. The limits on $\theta_{34}$ from low energy atmospheric neutrinos and solar neutrinos are comparable to and weaker than the MINOS upper limit, respectively. In Figure~\ref{fig:q34} solid and dashed curves correspond to sensitivity of DeepCore with effective volume $V_{\rm eff}^{\rm pes}$ and $V_{\rm eff}^{\rm opt}$; blue and green colors denote sensitivity for 2 and 3 bins of $\cos\theta_z$, respectively. Comparing left and right plots in Figure~\ref{fig:q34} shows that by increasing $\sin^22\theta_{24}$, sensitivity of DeepCore to $\theta_{34}$ increases, which is a result of more efficient $\bar{\nu}_\mu\to\bar{\nu}_\tau$ conversion. As can be seen, even with pessimistic assumption for effective volume and two bins of $\cos\theta_z$ (the blue solid curves), which is certainly realistic, it is possible to probe $\sin^2 2\theta_{34}$ values below the current upper limit. With pessimistic effective volume and two bins of $\cos\theta_z$, sensitivity of DeepCore is down to $\sin^2 2\theta_{34}\sim0.1$, which is a factor of six smaller than the current limit.

As an alternative way to obtain the limit on $\theta_{34}$, instead of fixing the value of $\theta_{24}$ in eq.~(\ref{eq:tildechi2}) it is possible to marginalize $\widetilde{\chi}^2$ with respect to $\theta_{24}$. However, it should be noticed that marginalizing with respect to $\theta_{24}$ over all possible values ({\it i.e.} $\sin^22\theta_{24}$ from 0 to 1) leads to no sensitivity to $\theta_{34}$, since the sensitivity to $\theta_{34}$ in our analysis is present when $\theta_{24}\neq0$. Thus, in principle we can marginalize with respect to $\theta_{24}$ over a range excluding zero, say $[\theta_{24}^{\rm min},\pi/4]$, where $\theta_{24}^{\rm min}>0$. However, it is straightforward to show that the limit on $\theta_{34}$ obtained from marginalizing $\widetilde{\chi}^2$ with respect to $\sin^22\theta_{24}$ over the range $[\sin^22\theta_{24}^{\rm min},1]$ is the same as the limit from eq.~(\ref{eq:tildechi2}) with $\sin^22\theta_{24}$ fixed to $\sin^22\theta_{24}^{\rm min}$.

\section{\label{sec:conclusion}Conclusion}

Several anomalies in neutrino oscillation experiments including reactor, Gallium, LSND and MiniBooNE anomalies, may hint at the presence of one (or more) almost sterile neutrino states in the mass range $\sim 1~{\rm eV}$ in addition to the established three active neutrino states. Simplest model accommodating a single sterile state is the so-called $3+1$ model that introduces four new mixing parameters: $\theta_{i4}$ and $\Delta m_{41}^2$ ($i=1,2,3$). Among the mixing angles, fits to reactor neutrino, LSND and MiniBooNE data indicate non-vanishing values for $\theta_{14}$ and $\theta_{24}$. For $\theta_{34}$ the only available information comes from comparing the rate of NC interaction between near and far detectors in MINOS experiment which constrain this mixing angle to $\sin^22\theta_{34}<0.59$ at 90\% C. L..

Active-sterile neutrino oscillations are enhanced by matter effects in propagation of the atmospheric neutrino flux through the Earth; for $\Delta m_{41}^2\sim1~{\rm eV}^2$ enhancement is maximal for neutrino energies of $E_\nu\sim$~TeV. This energy is well within the sensitivity range of large neutrino telescopes such as the IceCube detector at South Pole. The instrument detect neutrinos through $\mu$-tracks initiated by muon neutrinos and cascade events originating from electromagnetic and hadronic showers produced by neutrinos of all flavors. 

Specifically, nonzero values of $\theta_{14}$, $\theta_{24}$ and $\theta_{34}$ lead to enhancement of $\nu_e\to\nu_s$, $\bar{\nu}_\mu\to\bar{\nu}_s$ and $\bar{\nu}_\tau\to\bar{\nu}_s$ oscillations, respectively. As a result of absence of a significant number of $\bar{\nu}_\tau$ and small contribution of $\nu_e$ to the atmospheric neutrino flux, constraining $\theta_{14}$ and $\theta_{34}$ is challenging. In contrast constraining $\theta_{24}$ is feasible by measurement of the high statistics $\mu$-track events. Using this method the strongest upper limit on $\theta_{24}$ for masses below 1\,eV has been obtained by analyzing the IceCube-40 $\mu$-track data~\cite{Esmaili:2012nz}.  
 
In this paper we proposed a new method for the determination of mixing angle $\theta_{34}$  which quantifies $\nu_\tau-\nu_s$ mixing. The method is based on the fact that when both $\theta_{24}$ and $\theta_{34}$ are non-vanishing, the sterile state indirectly induces $\bar{\nu}_\mu\to\bar{\nu}_\tau$ conversion. Since the rates of $\bar{\nu}_\mu$ and $\bar{\nu}_\tau$-induced cascades in IceCube are different (the former generate cascades by NC interaction and the latter by both NC and CC interactions), $\bar{\nu}_\mu\to\bar{\nu}_\tau$ conversion distorts energy and zenith distributions of cascade events. We have shown that IceCube is sensitive to this distortion and by a few years of data-taking it is possible to probe $\theta_{34}$ values well below the current upper limit. 

In our analysis we included DeepCore part of IceCube detector which benefits from a lower energy threshold for detection of cascades that can be reconstructed with improved resolution in energy and direction. For the effective volume of DeepCore that is still being improved, we have considered two extreme cases such that the final effective volume will be between the pessimistic and optimistic estimates. We have shown that even with for the pessimistic estimate of effective volume, DeepCore will improve present constraints on $\theta_{34}$ angle by a factor of $\sim6$. 

We have also shown that nonzero values of $\theta_{24}$ and $\theta_{34}$ result into a unique signature in energy distribution of cascade events in DeepCore. While the presence of sterile neutrinos typically lead to a deficit in energy distribution of events through the $\bar{\nu}_\mu\to\bar{\nu}_s$ conversion, nonzero value of $\theta_{34}$ enhances $\bar{\nu}_\mu\to\bar{\nu}_\tau$ which leads to an increase in the number of cascade events with energy $\sim 3-5~{\rm TeV}~(\Delta m_{41}^2/{\rm eV}^2)$. This increase in number of cascade events is a direct manifestation of $\theta_{34}\neq0$, especially when combined with a corresponding deficit in number of $\mu$-tracks due to $\theta_{24}\neq0$.

\begin{acknowledgments} 
F.~H. acknowledges the support of the U.S. National Science Foundation-Office of Polar Programs, the U.S. National Science Foundation-Physics Division, the U.S. Department of Energy and the University of Wisconsin Alumni Research Foundation. O.~L.~G.~P. thanks ICTP and financial support from the funding grant 2012/16389-1, S\~ao Paulo Research Foundation (FAPESP). A.~E. acknowledges financial support from the funding grant 2010/13738-0, S\~ao Paulo Research Foundation (FAPESP). The authors thank CENAPAD and CCJDR for computing facilities.
\end{acknowledgments}


\begin{thebibliography}{30}

\bibitem{GonzalezGarcia:2007ib} 
  M.~C.~Gonzalez-Garcia and M.~Maltoni,
  Phys.\ Rept.\  {\bf 460}, 1 (2008)
  [arXiv:0704.1800 [hep-ph]].
  
  
\bibitem{Aguilar:2001ty} 
  A.~Aguilar-Arevalo {\it et al.}  [LSND Collaboration],
  Phys.\ Rev.\ D {\bf 64}, 112007 (2001)
  [hep-ex/0104049].
  
\bibitem{AguilarArevalo:2012va} 
  A.~A.~Aguilar-Arevalo {\it et al.}  [MiniBooNE Collaboration],
  arXiv:1207.4809 [hep-ex].
  
\bibitem{Aguilar-Arevalo:2013pmq} 
  A.~A.~Aguilar-Arevalo {\it et al.}  [MiniBooNE Collaboration],
  arXiv:1303.2588 [hep-ex].
  
  
\bibitem{Mueller:2011nm} 
  T.~.A.~Mueller, D.~Lhuillier, M.~Fallot, A.~Letourneau, S.~Cormon, M.~Fechner, L.~Giot and T.~Lasserre {\it et al.},
  Phys.\ Rev.\ C {\bf 83}, 054615 (2011)
  [arXiv:1101.2663 [hep-ex]].
  
\bibitem{Huber:2011wv} 
  P.~Huber,
  Phys.\ Rev.\ C {\bf 84}, 024617 (2011)
  [Erratum-ibid.\ C {\bf 85}, 029901 (2012)]
  [arXiv:1106.0687 [hep-ph]].
  
\bibitem{Kaether:2010ag} 
  F.~Kaether, W.~Hampel, G.~Heusser, J.~Kiko and T.~Kirsten,
  Phys.\ Lett.\ B {\bf 685}, 47 (2010)
  [arXiv:1001.2731 [hep-ex]].
  
\bibitem{Abdurashitov:2005tb} 
  J.~N.~Abdurashitov, V.~N.~Gavrin, S.~V.~Girin, V.~V.~Gorbachev, P.~P.~Gurkina, T.~V.~Ibragimova, A.~V.~Kalikhov and N.~G.~Khairnasov {\it et al.},
  Phys.\ Rev.\ C {\bf 73}, 045805 (2006)
  [nucl-ex/0512041].
  
  
\bibitem{Komatsu:2010fb} 
  E.~Komatsu {\it et al.}  [WMAP Collaboration],
  Astrophys.\ J.\ Suppl.\  {\bf 192}, 18 (2011)
  [arXiv:1001.4538 [astro-ph.CO]].
  
\bibitem{Seljak:2006bg} 
  U.~Seljak, A.~Slosar and P.~McDonald,
  JCAP {\bf 0610}, 014 (2006)
  [astro-ph/0604335].
  
\bibitem{GonzalezGarcia:2010un} 
  M.~C.~Gonzalez-Garcia, M.~Maltoni and J.~Salvado,
  JHEP {\bf 1008}, 117 (2010)
  [arXiv:1006.3795 [hep-ph]].
  
\bibitem{Archidiacono:2011gq} 
  M.~Archidiacono, E.~Calabrese and A.~Melchiorri,
  Phys.\ Rev.\ D {\bf 84}, 123008 (2011)
  [arXiv:1109.2767 [astro-ph.CO]].
  
\bibitem{Hamann:2010bk} 
  J.~Hamann, S.~Hannestad, G.~G.~Raffelt, I.~Tamborra and Y.~Y.~Y.~Wong,
  Phys.\ Rev.\ Lett.\  {\bf 105}, 181301 (2010)
  [arXiv:1006.5276 [hep-ph]].
  
\bibitem{Abazajian:2012ys} 
  K.~N.~Abazajian, M.~A.~Acero, S.~K.~Agarwalla, A.~A.~Aguilar-Arevalo, C.~H.~Albright, S.~Antusch, C.~A.~Arguelles and A.~B.~Balantekin {\it et al.},
  arXiv:1204.5379 [hep-ph].
  
  
\bibitem{Nunokawa:2003ep} 
  H.~Nunokawa, O.~L.~G.~Peres and R.~Zukanovich Funchal,
  Phys.\ Lett.\ B {\bf 562}, 279 (2003)
  [hep-ph/0302039].
  
\bibitem{Choubey:2007ji} 
  S.~Choubey,
  JHEP {\bf 0712}, 014 (2007)
  [arXiv:0709.1937 [hep-ph]].
  
  
\bibitem{Razzaque:2012tp} 
  S.~Razzaque and A.~Y.~Smirnov,
  Phys.\ Rev.\ D {\bf 85}, 093010 (2012)
  [arXiv:1203.5406 [hep-ph]].
  
\bibitem{Razzaque:2011ab} 
  S.~Razzaque and A.~Y.~Smirnov,
  JHEP {\bf 1107}, 084 (2011)
  [arXiv:1104.1390 [hep-ph]].
  
\bibitem{Barger:2011rc} 
  V.~Barger, Y.~Gao and D.~Marfatia,
  Phys.\ Rev.\ D {\bf 85}, 011302 (2012)
  [arXiv:1109.5748 [hep-ph]].
  
\bibitem{Halzen:2011yq} 
  F.~Halzen,
  arXiv:1111.0918 [hep-ph].
  
\bibitem{Esmaili:2012nz} 
  A.~Esmaili, F.~Halzen and O.~L.~G.~Peres,
  JCAP {\bf 1211}, 041 (2012)
  [arXiv:1206.6903 [hep-ph]].
  
  

\bibitem{deGouvea:2008nm}
  A.~de Gouvea and J.~Jenkins,
  Phys.\ Rev.\  D {\bf 78}, 053003 (2008)
  [arXiv:0804.3627 [hep-ph]].
  
  
\bibitem{Giunti:2012tn} 
  C.~Giunti, M.~Laveder, Y.~F.~Li, Q.~Y.~Liu and H.~W.~Long,
  Phys.\ Rev.\ D {\bf 86}, 113014 (2012)
  [arXiv:1210.5715 [hep-ph]].
  
\bibitem{Farzan:2001cj} 
  Y.~Farzan, O.~L.~G.~Peres and A.~Y.~.Smirnov,
  Nucl.\ Phys.\ B {\bf 612}, 59 (2001)
  [hep-ph/0105105].
  
\bibitem{Farzan:2002zq} 
  Y.~Farzan and A.~Y.~.Smirnov,
  Phys.\ Lett.\ B {\bf 557}, 224 (2003)
  [hep-ph/0211341].
  
\bibitem{deGouvea:2006gz} 
  A.~de Gouvea, J.~Jenkins and N.~Vasudevan,
  Phys.\ Rev.\ D {\bf 75}, 013003 (2007)
  [hep-ph/0608147].
  
\bibitem{Riis:2010zm} 
  A.~S.~Riis and S.~Hannestad,
  JCAP {\bf 1102}, 011 (2011)
  [arXiv:1008.1495 [astro-ph.CO]].
    
  
\bibitem{Formaggio:2011jg} 
  J.~A.~Formaggio and J.~Barrett,
  Phys.\ Lett.\ B {\bf 706}, 68 (2011)
  [arXiv:1105.1326 [nucl-ex]].
  
  
\bibitem{Kraus:2012he} 
  C.~Kraus, A.~Singer, K.~Valerius and C.~Weinheimer,
  arXiv:1210.4194 [hep-ex].
  
  
\bibitem{Esmaili:2012vg} 
  A.~Esmaili and O.~L.~G.~Peres,
  Phys.\ Rev.\ D {\bf 85}, 117301 (2012)
  [arXiv:1203.2632 [hep-ph]].
  

\bibitem{Adamson:2010wi} 
  P.~Adamson {\it et al.}  [MINOS Collaboration],
  Phys.\ Rev.\ D {\bf 81}, 052004 (2010)
  [arXiv:1001.0336 [hep-ex]].
    

\bibitem{Adamson:2011ku} 
  P.~Adamson {\it et al.}  [MINOS Collaboration],
  Phys.\ Rev.\ Lett.\  {\bf 107}, 011802 (2011)
  [arXiv:1104.3922 [hep-ex]].
  
\bibitem{Giunti:2011hn} 
  C.~Giunti and M.~Laveder,
  Phys.\ Rev.\ D {\bf 84}, 093006 (2011)
  [arXiv:1109.4033 [hep-ph]].
  
  
  
  
\bibitem{Peres:2000ic} 
  O.~L.~G.~Peres and A.~Y.~.Smirnov,
  Nucl.\ Phys.\ B {\bf 599}, 3 (2001)
  [hep-ph/0011054].
  
  
\bibitem{Kopp:2011qd} 
  J.~Kopp, M.~Maltoni and T.~Schwetz,
  Phys.\ Rev.\ Lett.\  {\bf 107}, 091801 (2011)
  [arXiv:1103.4570 [hep-ph]].
    
  
\bibitem{Donini:2007yf} 
  A.~Donini, M.~Maltoni, D.~Meloni, P.~Migliozzi and F.~Terranova,
  JHEP {\bf 0712}, 013 (2007)
  [arXiv:0704.0388 [hep-ph]].
  
\bibitem{Migliozzi:2011bj} 
  P.~Migliozzi and F.~Terranova,
  New J.\ Phys.\  {\bf 13}, 083016 (2011)
  [arXiv:1107.3018 [hep-ex]].
  
\bibitem{Donini:2001xy} 
  A.~Donini and D.~Meloni,
  Eur.\ Phys.\ J.\ C {\bf 22}, 179 (2001)
  [hep-ph/0105089].
  
\bibitem{Donini:2008wz} 
  A.~Donini, K.~-i.~Fuki, J.~Lopez-Pavon, D.~Meloni and O.~Yasuda,
  JHEP {\bf 0908}, 041 (2009)
  [arXiv:0812.3703 [hep-ph]].
  
\bibitem{Meloni:2010zr} 
  D.~Meloni, J.~Tang and W.~Winter,
  Phys.\ Rev.\ D {\bf 82}, 093008 (2010)
  [arXiv:1007.2419 [hep-ph]].
  
  
\bibitem{Maltoni:2007zf}  M.~Maltoni and T.~Schwetz,
  Phys.\ Rev.\ D {\bf 76}, 093005 (2007)
  [arXiv:0705.0107 [hep-ph]].
  
  
\bibitem{Maltoni:2001bc} 
  M.~Maltoni, T.~Schwetz and J.~W.~F.~Valle,
  Phys.\ Rev.\ D {\bf 65}, 093004 (2002)
  [hep-ph/0112103].
    
  
\bibitem{GonzalezGarcia:2012sz} 
  M.~C.~Gonzalez-Garcia, M.~Maltoni, J.~Salvado and T.~Schwetz,
  JHEP {\bf 1212}, 123 (2012)
  [arXiv:1209.3023 [hep-ph]].
  
  \bibitem{prem}  
A.~D.~Dziewonski and D.~L.~Anderson, Physics of the Earth and Planetary Interiors {\bf 25}, 297 (1981).    
  
\bibitem{Abbasi:2011ui} 
  R.~Abbasi {\it et al.}  [IceCube Collaboration],
  Phys.\ Rev.\ D {\bf 84}, 072001 (2011)
  [arXiv:1101.1692 [astro-ph.HE]].
  
  
\bibitem{Aartsen:2012uu} 
  M.~G.~Aartsen {\it et al.}  [IceCube Collaboration],
  Phys.\  Rev.\  Lett.\  {\bf 110}, 151105 (2013)
  [arXiv:1212.4760 [hep-ex]].
  
  
\bibitem{:2012zk} 
  [Chang Hyon Ha IceCube Collaboration],
  arXiv:1209.0698 [hep-ex];
  C.~H.~Ha [IceCube Collaboration],
  J.\ Phys.\ Conf.\ Ser.\  {\bf 375}, 052034 (2012)
  [arXiv:1201.0801 [hep-ex]].
  
\bibitem{Honda:2006qj} 
  M.~Honda, T.~Kajita, K.~Kasahara, S.~Midorikawa and T.~Sanuki,
  Phys.\ Rev.\ D {\bf 75}, 043006 (2007)
  [astro-ph/0611418];
  M.~Sajjad Athar, M.~Honda, T.~Kajita, K.~Kasahara and S.~Midorikawa,
  Phys.\ Lett.\ B {\bf 718}, 1375 (2013)
  [arXiv:1210.5154 [hep-ph]].
  
\bibitem{Collaboration:2011ym} 
  R.~Abbasi {\it et al.}  [IceCube Collaboration],
  Astropart.\ Phys.\  {\bf 35}, 615 (2012)
  [arXiv:1109.6096 [astro-ph.IM]].
       
    
\bibitem{icrc2009}
E. Middell {\it et al.}, ``{\it Improved Reconstruction of Cascade-like Events in IceCube}", accessible from \url{http://icecube.wisc.edu/reports/icrc2009}

\bibitem{thesis}
Stephanie Virginia Hickford, ``{\it A Cascade Analysis for the IceCube Neutrino Telescope}", accessible from \url{http://www2.phys.canterbury.ac.nz/~svh13/project.pdf}

\bibitem{Abbasi:2010ie} 
  R.~Abbasi {\it et al.}  [IceCube Collaboration],
  Phys.\ Rev.\ D {\bf 83}, 012001 (2011)
  [arXiv:1010.3980 [astro-ph.HE]].


\end{thebibliography}
\end{document}